\documentclass[app,amsmath,amssymb,reprint,superscriptaddress,floatfix]{revtex4-1}

\usepackage{graphicx}% Include figure files
\usepackage{dcolumn}% Align table columns on decimal point
\usepackage{bm}% bold math
\usepackage[utf8]{inputenc}
\usepackage[T1]{fontenc}
\usepackage{mathptmx}
\usepackage{siunitx}

\usepackage[normalem]{ulem}

\usepackage{hyperref}
\hypersetup{colorlinks=true, citecolor=blue, urlcolor=blue, linkcolor=red}
\usepackage{cleveref}
\Crefname{figure}{Figure}{}
\crefname{figure}{Fig.}{}

\begin{document}

% \preprint{AIP/123-QED}

\title{Silicon anisotropy in a bi-dimensional optomechanical cavity}
% Force line breaks with \\

\author{Cau\^e M. Kersul}
\altaffiliation{These two authors contributed equally} 
\affiliation{Photonics Research Center, University of Campinas, Campinas, SP, Brazil}
\affiliation{Applied Physics Department, Gleb Wataghin Physics Institute, University of Campinas, Campinas, SP, Brazil}

\author{Rodrigo Benevides}
%\altaffiliation[Now at ]{Department of Physics, ETH Zürich, 8093 Zürich, Switzerland} 
\altaffiliation{These two authors contributed equally} 
\affiliation{Photonics Research Center, University of Campinas, Campinas, SP, Brazil}
\affiliation{Applied Physics Department, Gleb Wataghin Physics Institute, University of Campinas, Campinas, SP, Brazil}
\affiliation{Kavli Institute of Nanoscience, Department of Quantum Nanoscience, Delft University of Technology, Delft, The Netherlands}

\author{Fl\'avio Moraes}
\affiliation{Photonics Research Center, University of Campinas, Campinas, SP, Brazil}
\affiliation{Applied Physics Department, Gleb Wataghin Physics Institute, University of Campinas, Campinas, SP, Brazil}

\author{Gabriel H. M. de Aguiar}
\affiliation{Photonics Research Center, University of Campinas, Campinas, SP, Brazil}
\affiliation{Applied Physics Department, Gleb Wataghin Physics Institute, University of Campinas, Campinas, SP, Brazil}

\author{Andreas Wallucks}
\affiliation{Kavli Institute of Nanoscience, Department of Quantum Nanoscience, Delft University of Technology, Delft, The Netherlands}

\author{Simon Gr\"{o}blacher}
\affiliation{Kavli Institute of Nanoscience, Department of Quantum Nanoscience, Delft University of Technology, Delft, The Netherlands}

\author{Gustavo S. Wiederhecker}
\affiliation{Photonics Research Center, University of Campinas, Campinas, SP, Brazil}
\affiliation{Quantum Electronics Department, Gleb Wataghin Physics Institute, University of Campinas, Campinas, SP, Brazil}

\author{Thiago P. Mayer Alegre}
\email{alegre@unicamp.br}
\affiliation{Photonics Research Center, University of Campinas, Campinas, SP, Brazil}
\affiliation{Applied Physics Department, Gleb Wataghin Physics Institute, University of Campinas, Campinas, SP, Brazil}

\date{\today}% It is always \today, today,

\begin{abstract}

In this work, we study the effects of mechanical anisotropy in a 2D optomechanical crystal geometry. We fabricate and measure devices with different orientations, showing the dependence of the mechanical spectrum and the optomechanical coupling with the relative angle of the device to the crystallography directions of silicon. Our results show that the device orientation strongly affects its mechanical band structure, which makes the devices more susceptible to fabrication imperfections. Finally, we show that our device is compatible with cryogenic measurements reaching ground state occupancy of 0.2 phonons at mK temperature.

\end{abstract}

\maketitle

In the last decade, optomechanical crystal cavities have been shown to confine light and mechanical motion in sub-wavelength modal volumes, leading to high optomechanical coupling rates~\cite{Leijssen2017,Balram2014, Chan2012} ($g_0$) and long-lived mechanical excitations~\cite{MayerAlegre:11,MacCabe2020}. Through the careful choice of geometry and material the photon-phonon interaction can be tailored, enabling applications not only in the classical realm, such as microwave phonon routing~\cite{Fang2016} and high frequency phonon sources~\cite{navarro2015self}, but also in the quantum domain, such as sideband ground-state cooling~\cite{Chan2011}, optomechanical quantum memories~\cite{wallucks2020quantum} and remote quantum state transfer in mechanical dual-rail encoded qubits~\cite{Fiaschi2021}.

%Since the early 2000s, integrated photonic crystals have been successfully used to confine light in sub-wavelength modal volumes to increase the interaction of light with matter. Purcell factor enhancement~\cite{Englund2005}, sensing applications~\cite{Momeni2009}, and channel drop filters~\cite{Akahane2003} are just a few examples that are favored by enhanced interaction. Moreover, when properly designed, a single device can confine long-lived mechanical excitations~\cite{Chan2012,MayerAlegre:11,MacCabe2020} and achieve high optomechanical coupling rate~\cite{Leijssen2017,Balram2014} ($g_0$), making it possible to explore not only classical aspects of photon-phonon interaction, such as microwave phonon routing in optomechanical circuits~\cite{Fang2016} and chaotic mechanical response~\cite{Navarro-Urrios2017}, but also quantum facets of it such as sideband ground-state cooling~\cite{Chan2011}, remote entanglement of mechanical oscillators~\cite{Riedinger2018} and the remote quantum state transfer using mechanical dual-rail encoded qubits~\cite{Fiaschi2021}.

Most of the optomechanical crystal devices used in quantum experiments are based on suspended quasi-one-dimensional (1D) beam structures. These devices combine large optical and mechanical quality factors with high $g_0$ in a simple design. However, such quasi-1D geometries present a fundamental drawback since they usually do not have good thermal dissipation and even the faintest light pulses can heat a device away from its mechanical ground state~\cite{MacCabe2020,Wallucks2020}. In contrast, quasi-two-dimensional (2D) structures~\cite{Benevides2017,MayerAlegre:11, Ren2020} have already been demonstrated to have impressive optical quality factors~\cite{Sekoguchi2014} and could have a much better thermal conductance. Nonetheless, few experimental works have focused on hypersonic (>GHz) quasi-2D optomechanical crystal~\cite{ ren2022topological, Florez2022} devices due to their more complex design and fabrication process. In this case, material crystalline anisotropy, known to affect the performance of quasi 1D-optomechanical devices~\cite{Jiang:19, burgwal2022enhanced, Balram2014}, becomes even more important. In this work, we study through simulation, fabrication, and measurements, the impacts of the mechanical anisotropy on a silicon-based cavity using the recently proposed design from Ref.~\onlinecite{Ren2020}. We then achieve high optomechanical coupling and low-phonon number occupancy necessary for quantum experiments. 

\begin{figure}[htp]
\centering
\includegraphics{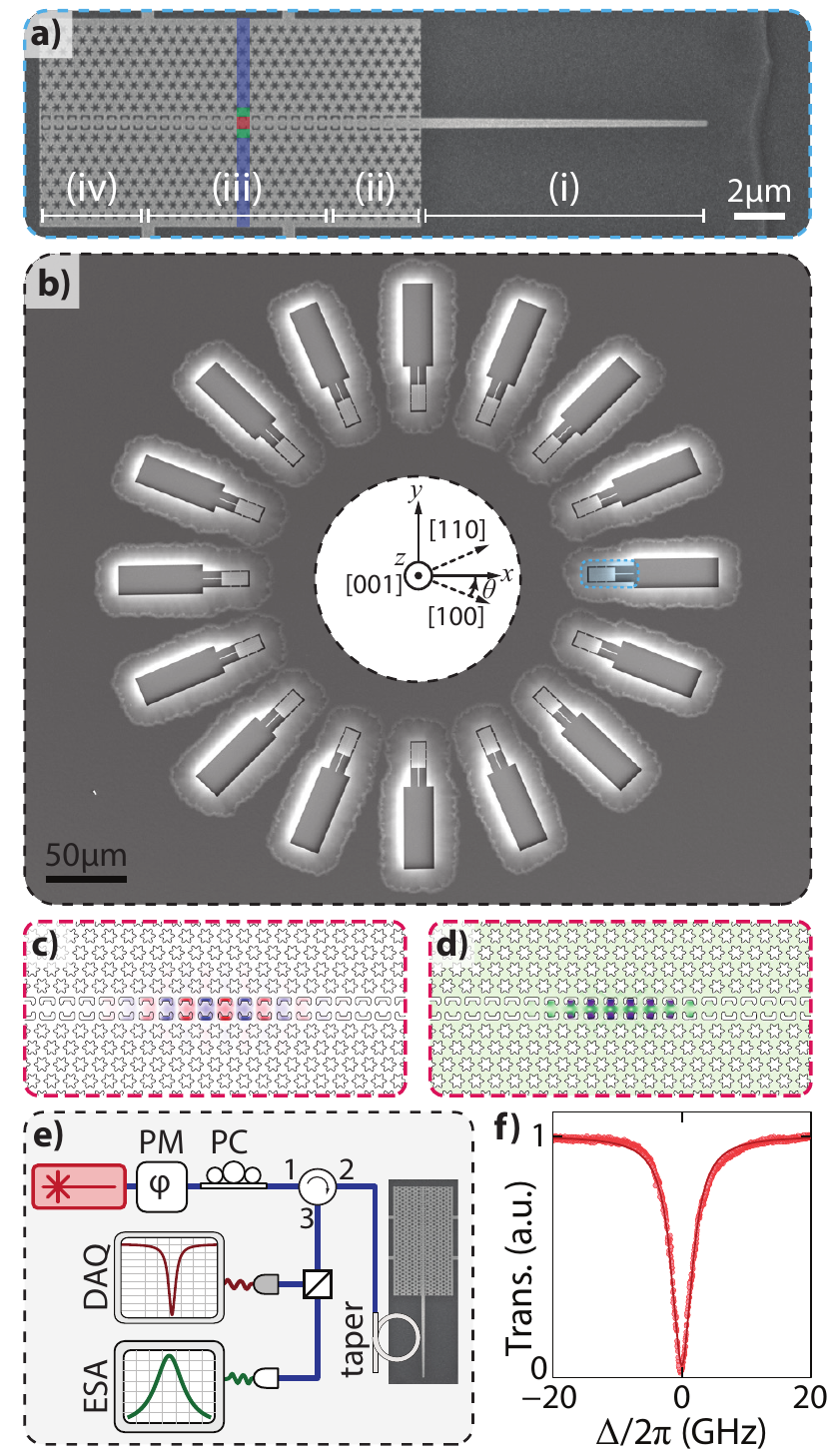}
\caption{\textbf{a)} and \textbf{b)}  Scanning Electron Microscopy (SEM) image of a single and a set of angled devices. In \textbf{a)} (i) indicates the tapered waveguide, (ii) the transition coupler, (iii) the defect, and (iv) the mirror region of our optomechanical crystal. \textbf{c)} and \textbf{d)}  Electrical ($E_y$) and mechanical ($|\mathbf{u}|$) fields of the confined optical and mechanical modes respectively. \textbf{e)} Experimental setup used in the measurements. \textbf{d)} Typical reflection signal of the optical resonance.} 
\label{fig1}
\end{figure}

The device's geometry consists of a quasi-1D cavity composed of two lines of C-shaped holes facing each other surrounded by a 2D triangular lattice of snowflake-shaped holes (\cref{fig1}~\textbf{a}). The snowflake structure allows for large optical and mechanical band gaps~\cite{safavinaeini2010, PhysRevLett.112.153603} that confine the modes within the C-shape region while still providing a good path for thermal dissipation compared to quasi-1D designs. The final devices, shown in Figs.~\ref{fig1}~\textbf{a-b}, are based on a traditional top-bottom fabrication process using a $\SI{220}{\nm}$ silicon-on-insulator (SOI) wafer. They are composed of: (i) a $\SI{10}{\um}$-long suspended tapered waveguide that efficiently couples~\cite{doi:10.1063/1.4826924} light from either a lensed fiber or a silica tapered fiber to a $\SI{360}{\nm}$ wide waveguide; (ii) a coupler region composed of eight transition cells, where the regular waveguide geometry is slowly morphed into the C-shape mirror unit cell, avoiding strong reflections due to impedance mismatching (see supplementary material for the specific geometry); (iii) a symmetrical defect region composed of 14 cells, whose dimensions are varied from the mirror unit cell at its edges to the defect unit cell at its center, confining both optical (\cref{fig1}~\textbf{c}) and mechanical (\cref{fig1}~\textbf{d}) modes; and finally (iv) a mirror region composed of eight mirror cells, placed at the end of the structure to avoid optical and mechanical leakage. The devices are fabricated at a given angle ($\theta$) defined as the counter-clockwise angle between the $[100]$ crystalline direction and the $x-$axis geometry while keeping the $z-$axis perpendicular to the wafer plane, which in our case is aligned to the $[001]$ direction. For a more detailed description of the fabricated design and of the fabrication process we refer the reader to section S1 of the Supplemental Material.  

\cref{fig1}~\textbf{e} shows the basic experimental setup used to measure both the optical and mechanical properties of the devices. It consists of a tunable laser, connected to an optical fiber circuit leading to a tapered fiber that couples light to the integrated tapered waveguides in our devices. Using a circulator before the tapered fiber we recover the reflection signal from our devices. With the aid of a data acquisition card and a slow photodetector, we measure the optical response of the cavities while slowly scanning the laser frequency, allowing us to characterize their optical response as a function of the detuning $\Delta$ between the laser and the optical mode frequency, as shown in \cref{fig1}~\textbf{f}. A fast detector attached to an electrical signal analyzer (ESA) is used to measure the Power Spectral Density (PSD) of the optical signal, showing the transduction of the mechanical spectrum due to the Brownian motion. The optomechanical coupling is then measured by comparing the intensity of the mechanical peaks with the intensity of a calibrated side-band introduced by a phase modulator, as proposed in ~\onlinecite{Gorodetksy:10}.

%\textbf{a)} Schematic of the optomechanical crystal based on the C-shape structure surrounded by the snowflake optical and mechanical shield.
%. The large rectangular holes around the snowflake shield are used to facilitate the undercut etch of the buried oxide layer.

\begin{figure*}[htp]
\centering
\includegraphics{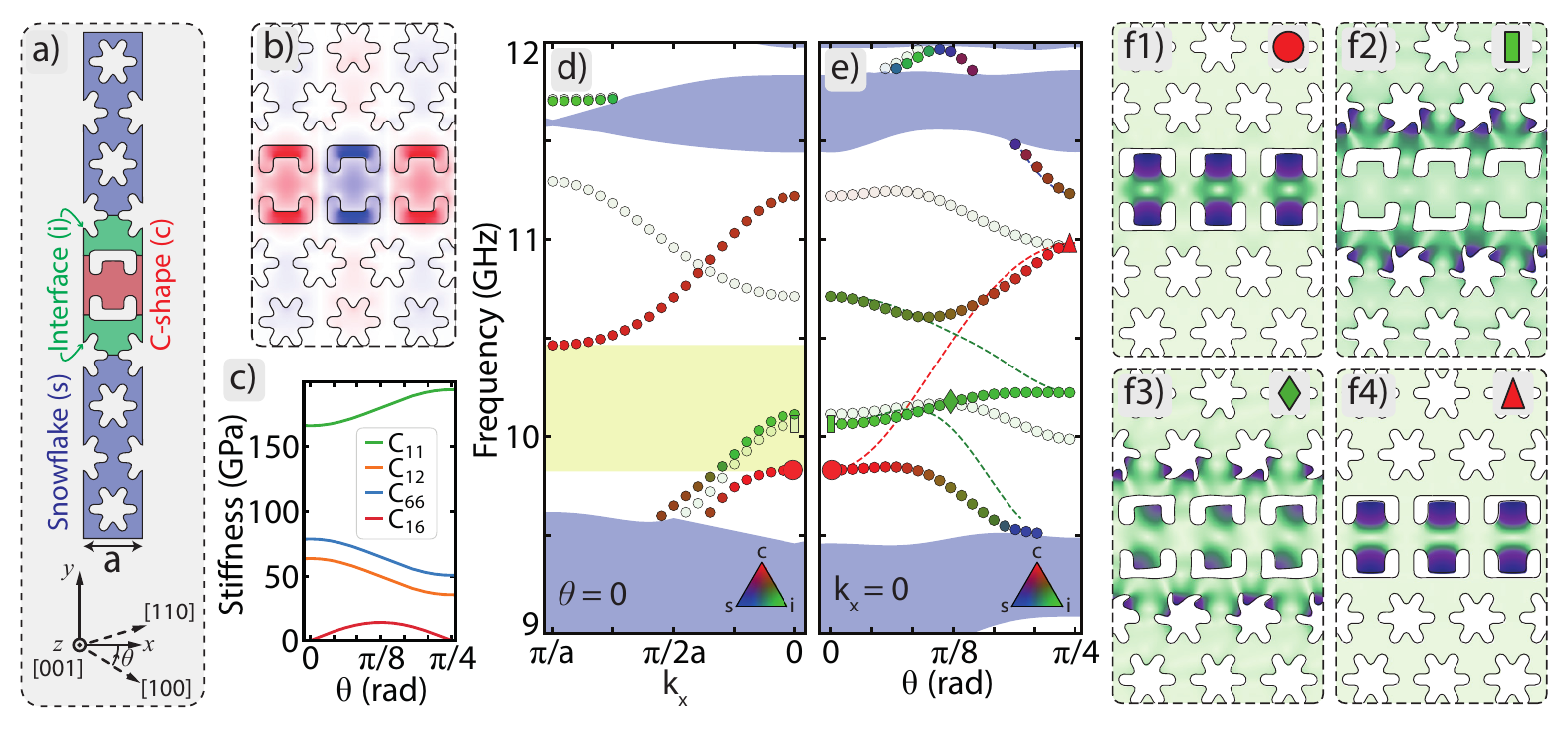}
\caption{\textbf{a)} Hybrid geometry unit cell separated in different regions. \textbf{b)} Electric field in the $y$ direction of the optical mode of interest. \textbf{c)} Different components of the silicon stiffness tensor. \textbf{d)} Mechanical band structure for $\theta = 0$. The colored circles indicate modes with $\sigma_y = +1$, while the transparent circles indicate modes with $\sigma_y = -1$. The RGB colors of the markers indicate the ratio of the displacement energy in each one of the regions of the unit cell, as indicated by the triangular legend. Notice that the $k_x$ axis is purposely inverted to approach modes at $k_x=0$ and $\theta = 0$ in  \textbf{d)} and \textbf{e)}. \textbf{e)} Mechanical band structure for $k_x = 0$ as a function of $\theta$. The colored circles indicate modes with $R_z^{\pi} = +1$ and the transparent circles indicate modes with $R_z^{\pi} = -1$. The dashed lines indicate the $R_z^{\pi} = +1$ modes when we force $C_{16} = 0$. The bigger markers in \textbf{d)} and \textbf{e)} indicate the modes presented at \textbf{f1)}-\textbf{f4)}. \textbf{f1)} The $(\sigma_x = +1, \sigma_y = +1)$ C-shape breathing mode for $k_x = 0$ and $\theta = 0$. \textbf{f2)} The $(\sigma_x = -1, \sigma_y = -1)$ interface mode for $k_x = 0$ and $\theta = 0$. \textbf{f3)} A $R_z^{\pi} = +1$ hybrid C-shape/interface mode for $k_x = 0$ and $\theta = \pi/8$. \textbf{f4)} The $(\sigma_x = +1, \sigma_y = +1)$ C-shape breathing mode for $k_x = 0$ and $\theta = \pi/4$.} 
\label{fig2}
\end{figure*}

In order to understand the fundamental impact of the mechanical anisotropy in the optomechanical coupling rate and mechanical mode confinement, we use Finite Element Method (FEM) simulations. Initially we explore how $\theta$ affects the band structure of a mechanical waveguide based on the 1D mirror unit cell of our devices (\cref{fig2}~\textbf{a}). These propagating modes are classified according to the symmetries of the waveguide, which are a composition of the symmetries arising from the unit-cell geometry and the silicon material properties. Understanding in detail such symmetries will allow us to better appreciate the behavior of the whole device, encompassing the mirror and the defect region.

The unit cell geometry in \cref{fig2}~\textbf{a} presents symmetry with respect to inversion about the $y-$ and $x-$axis ($\sigma_y$ and $\sigma_x$), nevertheless, Floquet periodicity over the $x-$axis is such that $\sigma_x$ symmetry is only valid for modes at the center or at the edges of the Brillouin zone ($k_x = 0$ or $k_x = \pi/a$). The components of the silicon stiffness tensor can also be classified according to the $\sigma_x$ and $\sigma_y$ symmetries. The basic components $C_{11}$, $C_{12}$ and $C_{66}$ present even symmetry, while the component $C_{16}$,  which couples compressive and shear stresses, is the only one that presents odd symmetry (\cref{fig2}~\textbf{c}). 

Due to the $C_{16}$ component, in general, the symmetry that is shared between the unit cell geometry and the material properties is a rotation of $\pi$ over the $z-$axis ($R_z^{\pi}$). Nevertheless, for $\theta = 0$ and $\theta = \pi/4$ the $x-$axis is aligned with the $[100]$ and $[110]$ crystallographic directions of silicon respectively, restoring the $\sigma_y$ and $\sigma_x$ symmetries as the $C_{16}$ component is null at these angles (\cref{fig2}~\textbf{c}).

In \cref{fig2}~\textbf{d} we present the mechanical band diagram for $\theta=0$ where the colored dots represent modes with even symmetry regarding $\sigma_y$ ($\sigma_y=+1$). The RGB color of the dots indicates the ratio of the displacement energy in each one of the highlighted regions in \cref{fig2}~\textbf{a}: the C-shape (red), the interface (green) and the snowflake (blue) regions. Transparent dots are modes with odd $\sigma_y$ symmetry ($\sigma_y=-1$), which do not couple with the even mechanical modes. When we consider only bands strongly confined at the C-shape region (red dots) we identify the $\sigma_y=+1$ apparent band gap highlighted by the yellow-shaded region in \cref{fig2}~\textbf{d}. This is not a true band gap, since a spurious $\sigma_y=+1$ interface band crosses it away from $k_x=0$ region.

The defect of our optomechanical crystal cavity was designed (see Supplemental Material S1) to confine the mode from the bottom C-shape band at $k_x=0$, which is shown in \cref{fig2}~\textbf{f1}. Despite the absence of a true phononic band gap, one can still achieve strong confinement as long as there is no significant energy exchange between the confined mode in question and the spurious interface modes~\cite{Eichenfield:09}. One of the factors limiting the coupling between those modes in the optomechanical cavity is the fact that at $k_x=0$ and $\theta=0$ the C-shape mode (\cref{fig2}~\textbf{f1}) is the only one that presents even symmetry for both $\sigma_x$ and $\sigma_y$. Additionally, the energy transfer from the C-shape to the spurious interface modes is inhibited by the much lower quality factor of the latter.

In \cref{fig2}~\textbf{e} we present the mechanical modes at the center of the Brillouin zone ($k_x=0$), and the dashed lines show the band structure with $C_{16}$ set to zero. In this artificial case, the C-shape mechanical mode (red dashed line) keeps its $\sigma_y$ symmetry and remains a pure C-shape mode. Its frequency monotonically increases with $\theta$, following the $C_{11}$ component related to pure compression, while it crosses modes with distinct symmetries located at the interface (green dashed lines). However, in the real case, the shear stress promoted by the $C_{16}$ component deforms the mechanical modes which are no longer symmetric under $\sigma_y$ and $\sigma_x$, but are still symmetric under $R_z$. The colored dots represent modes with even symmetry concerning $R_z^{\pi}$ ($R_z^{\pi}=+1$), while transparent dots present modes with $R_z^{\pi}=-1$. The color scheme is the same one used in the \cref{fig2}~\textbf{d} and represents the energy distribution in the geometry. As we go from $\theta = 0$ to $\theta = \pi/8$, the C-shape (\cref{fig2}~\textbf{f1}) and interface modes with the same $R_z$ symmetry, e.g. \cref{fig2}~\textbf{f2}, couple to each other leading to the large anti-crossing seen in \cref{fig2}~\textbf{e}. Consequently, there is no pure C-shape mode at $\theta=\pi/8$ but only mixed modes as the one shown in \cref{fig2}~\textbf{f3}. Furthermore, the anisotropy induced symmetry break and the strong delocalization away from the C-shape region decrease the overlap of such modes with the optical mode (\cref{fig2}~\textbf{b}), decreasing their optomechanical coupling.

As $\theta$ goes from $\pi/8$ to $\pi/4$ the value of the $C_{16}$ component decreases towards zero, the C-shape and interface modes become decoupled once again. At $\theta=\pi/4$, the original band gap does not exist anymore and a new band gap is formed between the C-shaped mode and the higher frequency snowflake modes (upper blue shaded region). There is also an interface mode very close in frequency to the C-shape mode and although, at $k_x = 0$, these two modes have distinct $\sigma_x$ symmetry, both have similar quality factors and share $\sigma_y$ symmetry, and, as such, fabrication imperfections are likely to couple them with each other. In addition, both bands present a flat dispersion near $k_x=0$ (see supplementary), making them even more susceptible to such imperfections \cite{PhysRevLett.112.153603}.   

\begin{figure*}[ht!]
\centering
\includegraphics{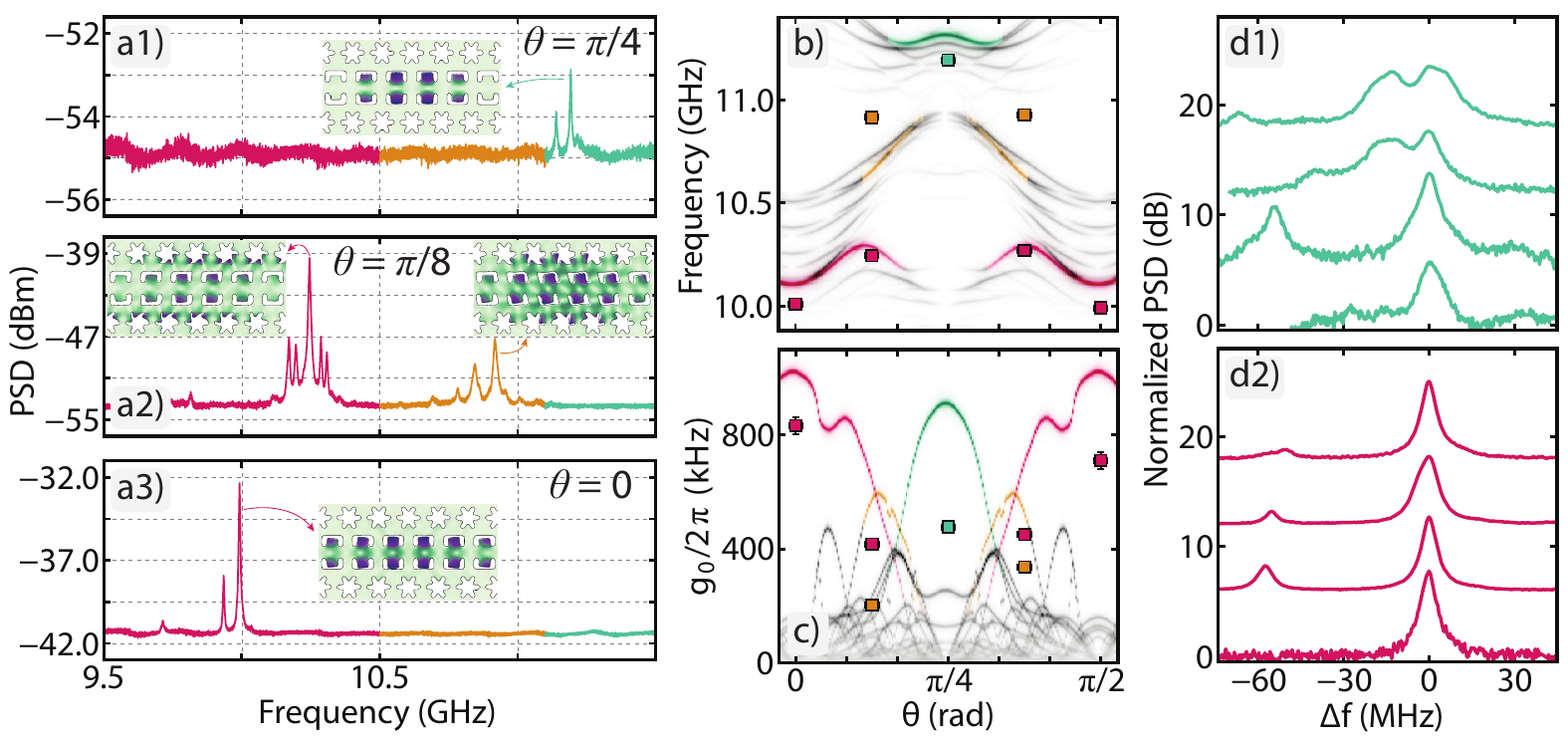}
\caption{\textbf{a1)} - \textbf{a3)} Power spectral density for devices at $\theta = \pi/4$, $\theta = \pi/8$ and $\theta = 0$. The insets present the displacement profiles of the indicated high $g_0$ mechanical modes. \textbf{b)} Density plot showing the simulated mechanical spectra as a function of $\theta$, for clarity only for $R_{z}^{\pi} = +1$ modes are shown. The highlighted modes are the ones with the highest simulated $g_0$ in each one of the frequency ranges. The square markers are experimental data, indicating the frequency of the higher $g_0$ modes in different devices. \textbf{c)} The solid lines indicate $g_0/2\pi$ for the modes highlighted in \textbf{b)}, the squares are the equivalent experimental data. \textbf{d1)} and \textbf{d2)} Mechanical spectra of 4 different devices for $\theta = \pi/4$ and $\theta = 0$ respectively. The spectra shown are normalized by the intensity of the PM and for visibility a common constant background is added to all spectra of the same type. The horizontal axis indicates the relative frequency with respect to the highest $g_0$ mode of each spectrum, while the different spectra are shifted in the vertical axis for better visualization.} 
\label{fig3}
\end{figure*}

Understanding the band structure upon which our optomechanical crystal cavity is based we then measured the devices shown in \cref{fig1}~\textbf{b}. The mechanical spectra of devices fabricated at $\theta =\pi/4$, $\pi/8$, and $0$ are shown in Figs.~\ref{fig3}~\textbf{a1-a3} respectively. Figs.~\textbf{b} and \textbf{c} present the results of the simulation for the whole device, encompassing both the mirror and the defect region. \cref{fig3}~\textbf{b} shows the frequencies of the simulated mechanical modes as a function of $\theta$, while Fig.~\ref{fig3} \textbf{c} shows the optomechanical coupling. The modes with the highest $g_0$ values in different frequency ranges are highlighted using the same color code as in Figs.~\ref{fig3}~\textbf{a1}-\textbf{a3}, defining mode branches. Both the experimental data and the simulations for the whole device follow the same behavior as in the waveguide simulations, where the frequency of the high $g_0$ C-shape modes increases as $\theta$ varies from $0$ to $\pi/4$, with a sudden frequency jump around $\theta = \pi/8$ due to the anti-crossing of C-shape and interface modes. As shown in Fig.~\ref{fig3} c), when passing through such anti-crossing the optomechanical coupling is transferred from one branch to the other. Inside the red branch, we also find a similar behavior with $g_0$ falling steeply due to a small anti-crossing with another interface mode in between $\theta = 0$ and $\theta = \pi/12$.

The displacement profiles shown in the insets of \cref{fig3}~\textbf{a1-a3} illustrate this picture. The C-shape mode at $\theta = 0$ is concentrated at the C-shape region, overlapping with the optical mode on the airgap. At $\theta = \pi/8$ hybridization occurs due to the anisotropy, and the mechanical displacement profile extends from the C-shape into the interface region, lowering the optomechanical overlap. Finally, for $\theta = \pi/4$ the displacement profile becomes again concentrated at the C-shape region. It is interesting to notice that the optomechanical coupling is higher for $\theta = 0$ than for $\theta = \pi/4$. This happens because the mode at $\theta = 0$ is more spread along the C-shape region than the mode at $\theta = \pi/4$\footnote{This is related to the flat dispersion of the C-shape mode at $\theta = \pi/4$ which leads to a high spectral density and consequently to modes very concentrated at few defect unit-cells.}, in such a way that the former has a better overlap with the optical mode.

The squares in Fig.~\ref{fig3} b) indicate the frequencies of the mechanical modes with the highest measured $g_0$~\cite{Gorodetksy:10} (see the S3 supplementary material for details on $g_0$ measurement). They agree within 1\% with the converged frequencies from the simulation. The optomechanical couplings on the other hand are not as well predicted as the frequencies, with the average highest measured $g_0$ varying between $70\%$ and $50\%$  of the simulated values for $\theta = 0$ and $\theta = \pi/4$ respectively. This indicates that the devices at $\theta = 0$ are more robust to fabrication imperfections than the ones at $\theta = \pi/4$. The comparison of different devices fabricated at the same orientation, shown in Figs.~\ref{fig3}~\textbf{d1} and \textbf{d2}, corroborates this analysis. For $\theta = 0$ the spectra of multiple devices, are similar to each other, usually presenting a mode with $g_0/2\pi$ between \SI{650}{kHz} and \SI{850}{kHz}, with a secondary mode at a frequency \SI{50}{MHz} lower, which corresponds quite well with the simulated second order C-shape mode. For $\theta = \pi/4$ the spectra are quite different among different devices and present multiple overlapping modes. Such lack of repeatability for $\theta = \pi/4$ can be attributed to the presence of flat C-shape and interface bands in the waveguide band structure, as discussed in Figs.~\ref{fig2}~\textbf{e}. 

For $\theta = \pi/8$ the comparison between simulation and experiment is more difficult as the estimated imprecision on the simulation of hybrid modes is larger (+4\%) than for the pure C-shape modes. Due to computational limitations, we cannot pinpoint the exact contribution of each interface mode to the measured modes, nevertheless, it is clear that multiple interface modes hybridize with the C-shape mode to form the two clusters of modes shown in \cref{fig3}~\textbf{a2}.

\begin{figure}
\centering
\includegraphics{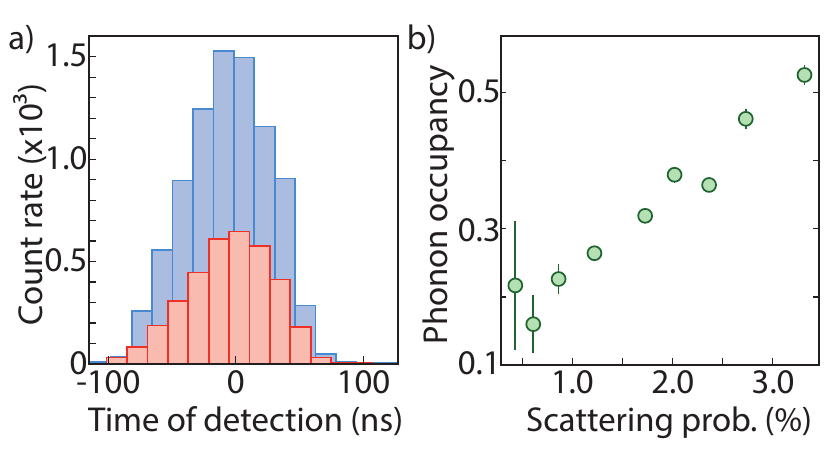}
\caption{ \textbf{a)} Typical asymmetry in the scattering count rates for the blue and red sidebands. \textbf{b)} Measured phonon occupancy as a function of pulse scattering probability.}
\label{fig4}
\end{figure}

As a demonstration of the compatibility of such devices with quantum optomechanics experiments, we measure below 1 phonon occupancies, placing them in a dilution fridge reaching $\SI{}{\milli \kelvin}$ base temperatures. The phonon occupancy ($\langle n_\text{pn}\rangle$) is measured using sideband asymmetry photon counting~\cite{PhysRevX.5.041002}. When a laser is fed in an optomechanical device, the interaction can be understood in terms of Stokes (anti-Stokes) processes in which one phonon is created (annihilated), while the frequency of one photon is decreased (increased) by one mechanical frequency ($\Omega/2\pi$). The likelihood of such scattering processes depends on the detuning, $\Delta$, of the laser relative to the optical resonance frequency. Stokes processes are maximum when the laser is blue-detuned by one mechanical frequency ($\Delta = +\Omega$), in contrast, anti-Stokes processes are maximum when the laser is red-detuned ($\Delta = -\Omega$).

To perform such an experiment we have used laser pulses that could be tuned to the blue and red sidebands. To avoid extra heating of the mechanical mode and guarantee that the initial state of the cavity is always the same, the laser pulses were tuned to be $\SI{100}{ns}$ wide with $\SI{100}{\us}$ period. The scattered photons were filtered, using $\SI{40}{MHz}$ wide Fabry-Pérot filters, and sent to single photon detectors. The filtering efficiency of the incident laser in both sidebands was on the order of $\SI{120}{dB}$.

\cref{fig4}~\textbf{a} shows an example of the scattering count rates obtained for the blue and red sidebands. As expected, we obtain different scattering rates for the two sidebands. In Fig~\ref{fig4}\textbf{b}, we show the resulting occupancy for different scattering probabilities (experimentally controlled via the pulse power), showing that $\langle n_\text{pn}\rangle$ falls as the probability is decreased. Moreover, with scattering probabilities below approximately $0.8\%$ (corresponding to approximately \SI{11}{nW}), we obtain an average occupation level lower than $0.2$ phonon in the system achieving a smaller occupation number for lower power, albeit with larger errors due to dark counts and leaking photons in the system.

% \section{Conclusion}

%In this work we have studied how the mechanical anisotropy affects the properties of bi-dimensional optomechanical crystals. The mechanical anisotropy was shown to couple confined defect modes with interface modes, decreasing the highest optomechanical coupling. This effect can be mitigated by choosing device orientations in which the device geometry shares the same symmetries as the mechanical anisotropy, $\theta = 0$ and $\theta = \pi/4$ in our case. We also demonstrate that our bidimensional optomechanical crystals are compatible with cryogenic measurements with phonon occupations below 0.2.

As the research on optomechanical devices progresses towards more complex materials, such as lithium niobate~\cite{Jiang:19} and gallium phosphide~\cite{stockill2022ultra, Schneider:19}, understanding the interplay between device geometry and material anisotropy will become increasingly important. Here we have studied how the silicon mechanical anisotropy affects the properties of our bi-dimensional optomechanical crystals. The mechanical anisotropy was shown to couple C-shape and interface modes, decreasing the overall optomechanical coupling. This effect can be mitigated by choosing orientations in which the device geometry shares the same symmetries as the device geometry, $\theta = 0$ and $\theta = \pi/4$ in our case. However, even at these angles, such couplings can still be relevant. In the future, we intend to investigate how to further suppress such couplings by the proper engineering of the interface region. For example, by stretching in the vertical direction the first line of snowflake holes it is possible to decrease the frequency of the interface modes changing their spectral distance from the c-shape mode at $\theta = \pi/4$. Finally, we also demonstrate that under passive cooling phonon occupations as low as 0.2 phonons can be measured, demonstrating, in conjunction with the thermal stability shown in ~\onlinecite{Ren2020}, its viability as a platform for quantum optomechanics experiments.

% In summary

% Despite the specificity of our design, we expect the same interplay between defect and interface modes to play a role in other geometries as long as they present the same hybrid geometry embedding .

\section*{SUPPLEMENTARY MATERIAL}
See Supplementary Material \textbf{S1} for details on C-shape geometry design and fabrication. \textbf{S2} provides a comprehensive description of the waveguide band structure for different crystalline orientations. Finally, in \textbf{S3} a detailed discussion of the optomechanical measurement and calibration is presented.

\section*{ACKNOWLEDGMENTS}
The authors would like to acknowledge assistance from the Kavli Nanolab Delft and CCSNano-UNICAMP with the micro-fabrication infrastructure. This work was supported by São Paulo Research Foundation (FAPESP) through grants 2019/01402-1, 2020/06348-2, 2020/00119-1, 2020/00100-9, 2022/07719-0, 2018/15580-6, 2018/15577-5, 2018/25339-4, Coordenação de Aperfeiçoamento de Pessoal de Nível Superior - Brasil (CAPES) (Finance Code 001), Financiadora de Estudos e Projetos (Finep), the European Research Council (ERC CoG Q-ECHOS, 101001005), and by the Netherlands Organization for Scientific Research (NWO/OCW), as part of the Frontiers of Nanoscience program, as well as through Vrij Programma (680-92-18-04).

\section*{DATA AVAILABILITY}
FEM and scripts files for generating each figure will be available at Ref.~\onlinecite{zenodo_cshape}. Additional data that support the findings of this study are available from the corresponding author upon reasonable request.

\bibliography{cshape-ref}% Produces the bibliography via BibTeX.

\begin{thebibliography}{28}%
\makeatletter
\providecommand \@ifxundefined [1]{%
 \@ifx{#1\undefined}
}%
\providecommand \@ifnum [1]{%
 \ifnum #1\expandafter \@firstoftwo
 \else \expandafter \@secondoftwo
 \fi
}%
\providecommand \@ifx [1]{%
 \ifx #1\expandafter \@firstoftwo
 \else \expandafter \@secondoftwo
 \fi
}%
\providecommand \natexlab [1]{#1}%
\providecommand \enquote  [1]{``#1''}%
\providecommand \bibnamefont  [1]{#1}%
\providecommand \bibfnamefont [1]{#1}%
\providecommand \citenamefont [1]{#1}%
\providecommand \href@noop [0]{\@secondoftwo}%
\providecommand \href [0]{\begingroup \@sanitize@url \@href}%
\providecommand \@href[1]{\@@startlink{#1}\@@href}%
\providecommand \@@href[1]{\endgroup#1\@@endlink}%
\providecommand \@sanitize@url [0]{\catcode `\\12\catcode `\$12\catcode
  `\&12\catcode `\#12\catcode `\^12\catcode `\_12\catcode `\%12\relax}%
\providecommand \@@startlink[1]{}%
\providecommand \@@endlink[0]{}%
\providecommand \url  [0]{\begingroup\@sanitize@url \@url }%
\providecommand \@url [1]{\endgroup\@href {#1}{\urlprefix }}%
\providecommand \urlprefix  [0]{URL }%
\providecommand \Eprint [0]{\href }%
\providecommand \doibase [0]{http://dx.doi.org/}%
\providecommand \selectlanguage [0]{\@gobble}%
\providecommand \bibinfo  [0]{\@secondoftwo}%
\providecommand \bibfield  [0]{\@secondoftwo}%
\providecommand \translation [1]{[#1]}%
\providecommand \BibitemOpen [0]{}%
\providecommand \bibitemStop [0]{}%
\providecommand \bibitemNoStop [0]{.\EOS\space}%
\providecommand \EOS [0]{\spacefactor3000\relax}%
\providecommand \BibitemShut  [1]{\csname bibitem#1\endcsname}%
\let\auto@bib@innerbib\@empty
%</preamble>
\bibitem [{\citenamefont {Leijssen}\ \emph {et~al.}(2017)\citenamefont
  {Leijssen}, \citenamefont {{La Gala}}, \citenamefont {Freisem}, \citenamefont
  {Muhonen},\ and\ \citenamefont {Verhagen}}]{Leijssen2017}%
  \BibitemOpen
  \bibfield  {author} {\bibinfo {author} {\bibfnamefont {R.}~\bibnamefont
  {Leijssen}}, \bibinfo {author} {\bibfnamefont {G.~R.}\ \bibnamefont {{La
  Gala}}}, \bibinfo {author} {\bibfnamefont {L.}~\bibnamefont {Freisem}},
  \bibinfo {author} {\bibfnamefont {J.~T.}\ \bibnamefont {Muhonen}}, \ and\
  \bibinfo {author} {\bibfnamefont {E.}~\bibnamefont {Verhagen}},\ }\href
  {\doibase 10.1038/ncomms16024} {\bibfield  {journal} {\bibinfo  {journal}
  {Nature Communications}\ }\textbf {\bibinfo {volume} {8}},\ \bibinfo {pages}
  {1} (\bibinfo {year} {2017})},\ \Eprint {http://arxiv.org/abs/1612.08072}
  {arXiv:1612.08072} \BibitemShut {NoStop}%
\bibitem [{\citenamefont {Balram}\ \emph {et~al.}(2014)\citenamefont {Balram},
  \citenamefont {Davanço}, \citenamefont {Lim}, \citenamefont {Song},\ and\
  \citenamefont {Srinivasan}}]{Balram2014}%
  \BibitemOpen
  \bibfield  {author} {\bibinfo {author} {\bibfnamefont {K.~C.}\ \bibnamefont
  {Balram}}, \bibinfo {author} {\bibfnamefont {M.}~\bibnamefont {Davanço}},
  \bibinfo {author} {\bibfnamefont {J.~Y.}\ \bibnamefont {Lim}}, \bibinfo
  {author} {\bibfnamefont {J.~D.}\ \bibnamefont {Song}}, \ and\ \bibinfo
  {author} {\bibfnamefont {K.}~\bibnamefont {Srinivasan}},\ }\href {\doibase
  10.1364/OPTICA.1.000414} {\bibfield  {journal} {\bibinfo  {journal} {Optica}\
  }\textbf {\bibinfo {volume} {1}},\ \bibinfo {pages} {414} (\bibinfo {year}
  {2014})}\BibitemShut {NoStop}%
\bibitem [{\citenamefont {Chan}\ \emph {et~al.}(2012)\citenamefont {Chan},
  \citenamefont {Safavi-Naeini}, \citenamefont {Hill}, \citenamefont
  {Meenehan},\ and\ \citenamefont {Painter}}]{Chan2012}%
  \BibitemOpen
  \bibfield  {author} {\bibinfo {author} {\bibfnamefont {J.}~\bibnamefont
  {Chan}}, \bibinfo {author} {\bibfnamefont {A.~H.}\ \bibnamefont
  {Safavi-Naeini}}, \bibinfo {author} {\bibfnamefont {J.~T.}\ \bibnamefont
  {Hill}}, \bibinfo {author} {\bibfnamefont {S.}~\bibnamefont {Meenehan}}, \
  and\ \bibinfo {author} {\bibfnamefont {O.}~\bibnamefont {Painter}},\ }\href
  {\doibase 10.1063/1.4747726} {\bibfield  {journal} {\bibinfo  {journal}
  {Applied Physics Letters}\ }\textbf {\bibinfo {volume} {101}},\ \bibinfo
  {pages} {081115} (\bibinfo {year} {2012})}\BibitemShut {NoStop}%
\bibitem [{\citenamefont {Alegre}\ \emph {et~al.}(2011)\citenamefont {Alegre},
  \citenamefont {Safavi-Naeini}, \citenamefont {Winger},\ and\ \citenamefont
  {Painter}}]{MayerAlegre:11}%
  \BibitemOpen
  \bibfield  {author} {\bibinfo {author} {\bibfnamefont {T.~P.~M.}\
  \bibnamefont {Alegre}}, \bibinfo {author} {\bibfnamefont {A.}~\bibnamefont
  {Safavi-Naeini}}, \bibinfo {author} {\bibfnamefont {M.}~\bibnamefont
  {Winger}}, \ and\ \bibinfo {author} {\bibfnamefont {O.}~\bibnamefont
  {Painter}},\ }\href {\doibase 10.1364/OE.19.005658} {\bibfield  {journal}
  {\bibinfo  {journal} {Opt. Express}\ }\textbf {\bibinfo {volume} {19}},\
  \bibinfo {pages} {5658} (\bibinfo {year} {2011})}\BibitemShut {NoStop}%
\bibitem [{\citenamefont {MacCabe}\ \emph {et~al.}(2020)\citenamefont
  {MacCabe}, \citenamefont {Ren}, \citenamefont {Luo}, \citenamefont {Cohen},
  \citenamefont {Zhou}, \citenamefont {Sipahigil}, \citenamefont
  {Mirhosseini},\ and\ \citenamefont {Painter}}]{MacCabe2020}%
  \BibitemOpen
  \bibfield  {author} {\bibinfo {author} {\bibfnamefont {G.~S.}\ \bibnamefont
  {MacCabe}}, \bibinfo {author} {\bibfnamefont {H.}~\bibnamefont {Ren}},
  \bibinfo {author} {\bibfnamefont {J.}~\bibnamefont {Luo}}, \bibinfo {author}
  {\bibfnamefont {J.~D.}\ \bibnamefont {Cohen}}, \bibinfo {author}
  {\bibfnamefont {H.}~\bibnamefont {Zhou}}, \bibinfo {author} {\bibfnamefont
  {A.}~\bibnamefont {Sipahigil}}, \bibinfo {author} {\bibfnamefont
  {M.}~\bibnamefont {Mirhosseini}}, \ and\ \bibinfo {author} {\bibfnamefont
  {O.}~\bibnamefont {Painter}},\ }\href {\doibase 10.1126/science.abc7312}
  {\bibfield  {journal} {\bibinfo  {journal} {Science}\ }\textbf {\bibinfo
  {volume} {370}},\ \bibinfo {pages} {840} (\bibinfo {year}
  {2020})}\BibitemShut {NoStop}%
\bibitem [{\citenamefont {Fang}\ \emph {et~al.}(2016)\citenamefont {Fang},
  \citenamefont {Matheny}, \citenamefont {Luan},\ and\ \citenamefont
  {Painter}}]{Fang2016}%
  \BibitemOpen
  \bibfield  {author} {\bibinfo {author} {\bibfnamefont {K.}~\bibnamefont
  {Fang}}, \bibinfo {author} {\bibfnamefont {M.~H.}\ \bibnamefont {Matheny}},
  \bibinfo {author} {\bibfnamefont {X.}~\bibnamefont {Luan}}, \ and\ \bibinfo
  {author} {\bibfnamefont {O.}~\bibnamefont {Painter}},\ }\href {\doibase
  10.1038/nphoton.2016.107} {\bibfield  {journal} {\bibinfo  {journal} {Nature
  Photonics}\ } (\bibinfo {year} {2016}),\
  10.1038/nphoton.2016.107}\BibitemShut {NoStop}%
\bibitem [{\citenamefont {Navarro-Urrios}\ \emph {et~al.}(2015)\citenamefont
  {Navarro-Urrios}, \citenamefont {Capuj}, \citenamefont {Gomis-Bresco},
  \citenamefont {Alzina}, \citenamefont {Pitanti}, \citenamefont {Griol},
  \citenamefont {Martínez},\ and\ \citenamefont {Torres}}]{navarro2015self}%
  \BibitemOpen
  \bibfield  {author} {\bibinfo {author} {\bibfnamefont {D.}~\bibnamefont
  {Navarro-Urrios}}, \bibinfo {author} {\bibfnamefont {N.~E.}\ \bibnamefont
  {Capuj}}, \bibinfo {author} {\bibfnamefont {J.}~\bibnamefont {Gomis-Bresco}},
  \bibinfo {author} {\bibfnamefont {F.}~\bibnamefont {Alzina}}, \bibinfo
  {author} {\bibfnamefont {A.}~\bibnamefont {Pitanti}}, \bibinfo {author}
  {\bibfnamefont {A.}~\bibnamefont {Griol}}, \bibinfo {author} {\bibfnamefont
  {A.}~\bibnamefont {Martínez}}, \ and\ \bibinfo {author} {\bibfnamefont
  {C.~M.~S.}\ \bibnamefont {Torres}},\ }\href {\doibase 10.1038/srep15733}
  {\bibfield  {journal} {\bibinfo  {journal} {Scientific Reports}\ }\textbf
  {\bibinfo {volume} {5}},\ \bibinfo {pages} {15733} (\bibinfo {year}
  {2015})}\BibitemShut {NoStop}%
\bibitem [{\citenamefont {Chan}\ \emph {et~al.}(2011)\citenamefont {Chan},
  \citenamefont {Alegre}, \citenamefont {Safavi-Naeini}, \citenamefont {Hill},
  \citenamefont {Krause}, \citenamefont {Gr{\"{o}}blacher}, \citenamefont
  {Aspelmeyer},\ and\ \citenamefont {Painter}}]{Chan2011}%
  \BibitemOpen
  \bibfield  {author} {\bibinfo {author} {\bibfnamefont {J.}~\bibnamefont
  {Chan}}, \bibinfo {author} {\bibfnamefont {T.~P.~M.}\ \bibnamefont {Alegre}},
  \bibinfo {author} {\bibfnamefont {A.~H.}\ \bibnamefont {Safavi-Naeini}},
  \bibinfo {author} {\bibfnamefont {J.~T.}\ \bibnamefont {Hill}}, \bibinfo
  {author} {\bibfnamefont {A.}~\bibnamefont {Krause}}, \bibinfo {author}
  {\bibfnamefont {S.}~\bibnamefont {Gr{\"{o}}blacher}}, \bibinfo {author}
  {\bibfnamefont {M.}~\bibnamefont {Aspelmeyer}}, \ and\ \bibinfo {author}
  {\bibfnamefont {O.}~\bibnamefont {Painter}},\ }\href {\doibase
  10.1038/nature10461} {\bibfield  {journal} {\bibinfo  {journal} {Nature}\
  }\textbf {\bibinfo {volume} {478}},\ \bibinfo {pages} {89} (\bibinfo {year}
  {2011})}\BibitemShut {NoStop}%
\bibitem [{\citenamefont {Wallucks}\ \emph
  {et~al.}(2020{\natexlab{a}})\citenamefont {Wallucks}, \citenamefont
  {Marinković}, \citenamefont {Hensen}, \citenamefont {Stockill},\ and\
  \citenamefont {Gröblacher}}]{wallucks2020quantum}%
  \BibitemOpen
  \bibfield  {author} {\bibinfo {author} {\bibfnamefont {A.}~\bibnamefont
  {Wallucks}}, \bibinfo {author} {\bibfnamefont {I.}~\bibnamefont
  {Marinković}}, \bibinfo {author} {\bibfnamefont {B.}~\bibnamefont {Hensen}},
  \bibinfo {author} {\bibfnamefont {R.}~\bibnamefont {Stockill}}, \ and\
  \bibinfo {author} {\bibfnamefont {S.}~\bibnamefont {Gröblacher}},\ }\href
  {\doibase 10.1038/s41567-020-0891-z} {\bibfield  {journal} {\bibinfo
  {journal} {Nature Physics}\ }\textbf {\bibinfo {volume} {16}},\ \bibinfo
  {pages} {772} (\bibinfo {year} {2020}{\natexlab{a}})}\BibitemShut {NoStop}%
\bibitem [{\citenamefont {Fiaschi}\ \emph {et~al.}(2021)\citenamefont
  {Fiaschi}, \citenamefont {Hensen}, \citenamefont {Wallucks}, \citenamefont
  {Benevides}, \citenamefont {Li}, \citenamefont {Alegre},\ and\ \citenamefont
  {Gröblacher}}]{Fiaschi2021}%
  \BibitemOpen
  \bibfield  {author} {\bibinfo {author} {\bibfnamefont {N.}~\bibnamefont
  {Fiaschi}}, \bibinfo {author} {\bibfnamefont {B.}~\bibnamefont {Hensen}},
  \bibinfo {author} {\bibfnamefont {A.}~\bibnamefont {Wallucks}}, \bibinfo
  {author} {\bibfnamefont {R.}~\bibnamefont {Benevides}}, \bibinfo {author}
  {\bibfnamefont {J.}~\bibnamefont {Li}}, \bibinfo {author} {\bibfnamefont
  {T.~P.~M.}\ \bibnamefont {Alegre}}, \ and\ \bibinfo {author} {\bibfnamefont
  {S.}~\bibnamefont {Gröblacher}},\ }\href {\doibase
  10.1038/s41566-021-00866-z} {\bibfield  {journal} {\bibinfo  {journal}
  {Nature Photonics}\ }\textbf {\bibinfo {volume} {15}},\ \bibinfo {pages}
  {817} (\bibinfo {year} {2021})}\BibitemShut {NoStop}%
\bibitem [{\citenamefont {Wallucks}\ \emph
  {et~al.}(2020{\natexlab{b}})\citenamefont {Wallucks}, \citenamefont
  {Marinkovi{\'{c}}}, \citenamefont {Hensen}, \citenamefont {Stockill},\ and\
  \citenamefont {Gr{\"{o}}blacher}}]{Wallucks2020}%
  \BibitemOpen
  \bibfield  {author} {\bibinfo {author} {\bibfnamefont {A.}~\bibnamefont
  {Wallucks}}, \bibinfo {author} {\bibfnamefont {I.}~\bibnamefont
  {Marinkovi{\'{c}}}}, \bibinfo {author} {\bibfnamefont {B.}~\bibnamefont
  {Hensen}}, \bibinfo {author} {\bibfnamefont {R.}~\bibnamefont {Stockill}}, \
  and\ \bibinfo {author} {\bibfnamefont {S.}~\bibnamefont {Gr{\"{o}}blacher}},\
  }\href {\doibase 10.1038/s41567-020-0891-z} {\bibfield  {journal} {\bibinfo
  {journal} {Nature Physics}\ }\textbf {\bibinfo {volume} {16}},\ \bibinfo
  {pages} {772} (\bibinfo {year} {2020}{\natexlab{b}})},\ \Eprint
  {http://arxiv.org/abs/1910.07409} {arXiv:1910.07409} \BibitemShut {NoStop}%
\bibitem [{\citenamefont {Benevides}\ \emph {et~al.}(2017)\citenamefont
  {Benevides}, \citenamefont {Santos}, \citenamefont {Luiz}, \citenamefont
  {Wiederhecker},\ and\ \citenamefont {Alegre}}]{Benevides2017}%
  \BibitemOpen
  \bibfield  {author} {\bibinfo {author} {\bibfnamefont {R.}~\bibnamefont
  {Benevides}}, \bibinfo {author} {\bibfnamefont {F.~G.~S.}\ \bibnamefont
  {Santos}}, \bibinfo {author} {\bibfnamefont {G.~O.}\ \bibnamefont {Luiz}},
  \bibinfo {author} {\bibfnamefont {G.~S.}\ \bibnamefont {Wiederhecker}}, \
  and\ \bibinfo {author} {\bibfnamefont {T.~P.~M.}\ \bibnamefont {Alegre}},\
  }\href {\doibase 10.1038/s41598-017-02515-4} {\bibfield  {journal} {\bibinfo
  {journal} {Scientific Reports}\ }\textbf {\bibinfo {volume} {7}},\ \bibinfo
  {pages} {2491} (\bibinfo {year} {2017})},\ \Eprint
  {http://arxiv.org/abs/1701.03410} {arXiv:1701.03410} \BibitemShut {NoStop}%
\bibitem [{\citenamefont {Ren}\ \emph {et~al.}(2020)\citenamefont {Ren},
  \citenamefont {Matheny}, \citenamefont {MacCabe}, \citenamefont {Luo},
  \citenamefont {Pfeifer}, \citenamefont {Mirhosseini},\ and\ \citenamefont
  {Painter}}]{Ren2020}%
  \BibitemOpen
  \bibfield  {author} {\bibinfo {author} {\bibfnamefont {H.}~\bibnamefont
  {Ren}}, \bibinfo {author} {\bibfnamefont {M.~H.}\ \bibnamefont {Matheny}},
  \bibinfo {author} {\bibfnamefont {G.~S.}\ \bibnamefont {MacCabe}}, \bibinfo
  {author} {\bibfnamefont {J.}~\bibnamefont {Luo}}, \bibinfo {author}
  {\bibfnamefont {H.}~\bibnamefont {Pfeifer}}, \bibinfo {author} {\bibfnamefont
  {M.}~\bibnamefont {Mirhosseini}}, \ and\ \bibinfo {author} {\bibfnamefont
  {O.}~\bibnamefont {Painter}},\ }\href {\doibase 10.1038/s41467-020-17182-9}
  {\bibfield  {journal} {\bibinfo  {journal} {Nature Communications}\ }\textbf
  {\bibinfo {volume} {11}},\ \bibinfo {pages} {3373} (\bibinfo {year}
  {2020})},\ \Eprint {http://arxiv.org/abs/1910.02873} {arXiv:1910.02873}
  \BibitemShut {NoStop}%
\bibitem [{\citenamefont {Sekoguchi}\ \emph {et~al.}(2014)\citenamefont
  {Sekoguchi}, \citenamefont {Takahashi}, \citenamefont {Asano},\ and\
  \citenamefont {Noda}}]{Sekoguchi2014}%
  \BibitemOpen
  \bibfield  {author} {\bibinfo {author} {\bibfnamefont {H.}~\bibnamefont
  {Sekoguchi}}, \bibinfo {author} {\bibfnamefont {Y.}~\bibnamefont
  {Takahashi}}, \bibinfo {author} {\bibfnamefont {T.}~\bibnamefont {Asano}}, \
  and\ \bibinfo {author} {\bibfnamefont {S.}~\bibnamefont {Noda}},\ }\href
  {\doibase 10.1364/OE.22.000916} {\bibfield  {journal} {\bibinfo  {journal}
  {Optics Express}\ }\textbf {\bibinfo {volume} {22}},\ \bibinfo {pages} {916}
  (\bibinfo {year} {2014})}\BibitemShut {NoStop}%
\bibitem [{\citenamefont {Ren}\ \emph {et~al.}(2022)\citenamefont {Ren},
  \citenamefont {Shah}, \citenamefont {Pfeifer}, \citenamefont {Brendel},
  \citenamefont {Peano}, \citenamefont {Marquardt},\ and\ \citenamefont
  {Painter}}]{ren2022topological}%
  \BibitemOpen
  \bibfield  {author} {\bibinfo {author} {\bibfnamefont {H.}~\bibnamefont
  {Ren}}, \bibinfo {author} {\bibfnamefont {T.}~\bibnamefont {Shah}}, \bibinfo
  {author} {\bibfnamefont {H.}~\bibnamefont {Pfeifer}}, \bibinfo {author}
  {\bibfnamefont {C.}~\bibnamefont {Brendel}}, \bibinfo {author} {\bibfnamefont
  {V.}~\bibnamefont {Peano}}, \bibinfo {author} {\bibfnamefont
  {F.}~\bibnamefont {Marquardt}}, \ and\ \bibinfo {author} {\bibfnamefont
  {O.}~\bibnamefont {Painter}},\ }\href {\doibase 10.1038/s41467-022-30941-0}
  {\bibfield  {journal} {\bibinfo  {journal} {Nature Communications}\ }\textbf
  {\bibinfo {volume} {13}},\ \bibinfo {pages} {3476} (\bibinfo {year}
  {2022})}\BibitemShut {NoStop}%
\bibitem [{\citenamefont {Florez}\ \emph {et~al.}(2022)\citenamefont {Florez},
  \citenamefont {Arregui}, \citenamefont {Albrechtsen}, \citenamefont {Ng},
  \citenamefont {Gomis-Bresco}, \citenamefont {Stobbe}, \citenamefont
  {Sotomayor-Torres},\ and\ \citenamefont {García}}]{Florez2022}%
  \BibitemOpen
  \bibfield  {author} {\bibinfo {author} {\bibfnamefont {O.}~\bibnamefont
  {Florez}}, \bibinfo {author} {\bibfnamefont {G.}~\bibnamefont {Arregui}},
  \bibinfo {author} {\bibfnamefont {M.}~\bibnamefont {Albrechtsen}}, \bibinfo
  {author} {\bibfnamefont {R.~C.}\ \bibnamefont {Ng}}, \bibinfo {author}
  {\bibfnamefont {J.}~\bibnamefont {Gomis-Bresco}}, \bibinfo {author}
  {\bibfnamefont {S.}~\bibnamefont {Stobbe}}, \bibinfo {author} {\bibfnamefont
  {C.~M.}\ \bibnamefont {Sotomayor-Torres}}, \ and\ \bibinfo {author}
  {\bibfnamefont {P.~D.}\ \bibnamefont {García}},\ }\href {\doibase
  10.1038/s41565-022-01178-1} {\bibfield  {journal} {\bibinfo  {journal}
  {Nature Nanotechnology}\ }\textbf {\bibinfo {volume} {17}},\ \bibinfo {pages}
  {947} (\bibinfo {year} {2022})}\BibitemShut {NoStop}%
\bibitem [{\citenamefont {Jiang}\ \emph {et~al.}(2019)\citenamefont {Jiang},
  \citenamefont {Patel}, \citenamefont {Mayor}, \citenamefont {McKenna},
  \citenamefont {Arrangoiz-Arriola}, \citenamefont {Sarabalis}, \citenamefont
  {Witmer}, \citenamefont {Laer},\ and\ \citenamefont
  {Safavi-Naeini}}]{Jiang:19}%
  \BibitemOpen
  \bibfield  {author} {\bibinfo {author} {\bibfnamefont {W.}~\bibnamefont
  {Jiang}}, \bibinfo {author} {\bibfnamefont {R.~N.}\ \bibnamefont {Patel}},
  \bibinfo {author} {\bibfnamefont {F.~M.}\ \bibnamefont {Mayor}}, \bibinfo
  {author} {\bibfnamefont {T.~P.}\ \bibnamefont {McKenna}}, \bibinfo {author}
  {\bibfnamefont {P.}~\bibnamefont {Arrangoiz-Arriola}}, \bibinfo {author}
  {\bibfnamefont {C.~J.}\ \bibnamefont {Sarabalis}}, \bibinfo {author}
  {\bibfnamefont {J.~D.}\ \bibnamefont {Witmer}}, \bibinfo {author}
  {\bibfnamefont {R.~V.}\ \bibnamefont {Laer}}, \ and\ \bibinfo {author}
  {\bibfnamefont {A.~H.}\ \bibnamefont {Safavi-Naeini}},\ }\href {\doibase
  10.1364/OPTICA.6.000845} {\bibfield  {journal} {\bibinfo  {journal} {Optica}\
  }\textbf {\bibinfo {volume} {6}},\ \bibinfo {pages} {845} (\bibinfo {year}
  {2019})}\BibitemShut {NoStop}%
\bibitem [{\citenamefont {Burgwal}\ and\ \citenamefont
  {Verhagen}(2022)}]{burgwal2022enhanced}%
  \BibitemOpen
  \bibfield  {author} {\bibinfo {author} {\bibfnamefont {R.}~\bibnamefont
  {Burgwal}}\ and\ \bibinfo {author} {\bibfnamefont {E.}~\bibnamefont
  {Verhagen}},\ }\href {http://arxiv.org/abs/2207.11114} {\bibfield  {journal}
  {\bibinfo  {journal} {arXiv preprint arXiv:2207.11114}\ } (\bibinfo {year}
  {2022})}\BibitemShut {NoStop}%
\bibitem [{\citenamefont {Safavi-Naeini}\ and\ \citenamefont
  {Painter}(2010)}]{safavinaeini2010}%
  \BibitemOpen
  \bibfield  {author} {\bibinfo {author} {\bibfnamefont {A.~H.}\ \bibnamefont
  {Safavi-Naeini}}\ and\ \bibinfo {author} {\bibfnamefont {O.}~\bibnamefont
  {Painter}},\ }\href {\doibase 10.1364/OE.18.014926} {\bibfield  {journal}
  {\bibinfo  {journal} {Optics Express}\ }\textbf {\bibinfo {volume} {18}},\
  \bibinfo {pages} {14926} (\bibinfo {year} {2010})}\BibitemShut {NoStop}%
\bibitem [{\citenamefont {Safavi-Naeini}\ \emph {et~al.}(2014)\citenamefont
  {Safavi-Naeini}, \citenamefont {Hill}, \citenamefont {Meenehan},
  \citenamefont {Chan}, \citenamefont {Gr\"oblacher},\ and\ \citenamefont
  {Painter}}]{PhysRevLett.112.153603}%
  \BibitemOpen
  \bibfield  {author} {\bibinfo {author} {\bibfnamefont {A.~H.}\ \bibnamefont
  {Safavi-Naeini}}, \bibinfo {author} {\bibfnamefont {J.~T.}\ \bibnamefont
  {Hill}}, \bibinfo {author} {\bibfnamefont {S.}~\bibnamefont {Meenehan}},
  \bibinfo {author} {\bibfnamefont {J.}~\bibnamefont {Chan}}, \bibinfo {author}
  {\bibfnamefont {S.}~\bibnamefont {Gr\"oblacher}}, \ and\ \bibinfo {author}
  {\bibfnamefont {O.}~\bibnamefont {Painter}},\ }\href {\doibase
  10.1103/PhysRevLett.112.153603} {\bibfield  {journal} {\bibinfo  {journal}
  {Phys. Rev. Lett.}\ }\textbf {\bibinfo {volume} {112}},\ \bibinfo {pages}
  {153603} (\bibinfo {year} {2014})}\BibitemShut {NoStop}%
\bibitem [{\citenamefont {Gröblacher}\ \emph {et~al.}(2013)\citenamefont
  {Gröblacher}, \citenamefont {Hill}, \citenamefont {Safavi-Naeini},
  \citenamefont {Chan},\ and\ \citenamefont {Painter}}]{doi:10.1063/1.4826924}%
  \BibitemOpen
  \bibfield  {author} {\bibinfo {author} {\bibfnamefont {S.}~\bibnamefont
  {Gröblacher}}, \bibinfo {author} {\bibfnamefont {J.~T.}\ \bibnamefont
  {Hill}}, \bibinfo {author} {\bibfnamefont {A.~H.}\ \bibnamefont
  {Safavi-Naeini}}, \bibinfo {author} {\bibfnamefont {J.}~\bibnamefont {Chan}},
  \ and\ \bibinfo {author} {\bibfnamefont {O.}~\bibnamefont {Painter}},\ }\href
  {\doibase 10.1063/1.4826924} {\bibfield  {journal} {\bibinfo  {journal}
  {Applied Physics Letters}\ }\textbf {\bibinfo {volume} {103}},\ \bibinfo
  {pages} {181104} (\bibinfo {year} {2013})},\ \Eprint
  {http://arxiv.org/abs/https://doi.org/10.1063/1.4826924}
  {https://doi.org/10.1063/1.4826924} \BibitemShut {NoStop}%
\bibitem [{\citenamefont {Gorodetksy}\ \emph {et~al.}(2010)\citenamefont
  {Gorodetksy}, \citenamefont {Schliesser}, \citenamefont {Anetsberger},
  \citenamefont {Deleglise},\ and\ \citenamefont {Kippenberg}}]{Gorodetksy:10}%
  \BibitemOpen
  \bibfield  {author} {\bibinfo {author} {\bibfnamefont {M.~L.}\ \bibnamefont
  {Gorodetksy}}, \bibinfo {author} {\bibfnamefont {A.}~\bibnamefont
  {Schliesser}}, \bibinfo {author} {\bibfnamefont {G.}~\bibnamefont
  {Anetsberger}}, \bibinfo {author} {\bibfnamefont {S.}~\bibnamefont
  {Deleglise}}, \ and\ \bibinfo {author} {\bibfnamefont {T.~J.}\ \bibnamefont
  {Kippenberg}},\ }\href {\doibase 10.1364/OE.18.023236} {\bibfield  {journal}
  {\bibinfo  {journal} {Opt. Express}\ }\textbf {\bibinfo {volume} {18}},\
  \bibinfo {pages} {23236} (\bibinfo {year} {2010})}\BibitemShut {NoStop}%
\bibitem [{\citenamefont {Eichenfield}\ \emph {et~al.}(2009)\citenamefont
  {Eichenfield}, \citenamefont {Chan}, \citenamefont {Safavi-Naeini},
  \citenamefont {Vahala},\ and\ \citenamefont {Painter}}]{Eichenfield:09}%
  \BibitemOpen
  \bibfield  {author} {\bibinfo {author} {\bibfnamefont {M.}~\bibnamefont
  {Eichenfield}}, \bibinfo {author} {\bibfnamefont {J.}~\bibnamefont {Chan}},
  \bibinfo {author} {\bibfnamefont {A.~H.}\ \bibnamefont {Safavi-Naeini}},
  \bibinfo {author} {\bibfnamefont {K.~J.}\ \bibnamefont {Vahala}}, \ and\
  \bibinfo {author} {\bibfnamefont {O.}~\bibnamefont {Painter}},\ }\href
  {\doibase 10.1364/OE.17.020078} {\bibfield  {journal} {\bibinfo  {journal}
  {Opt. Express}\ }\textbf {\bibinfo {volume} {17}},\ \bibinfo {pages} {20078}
  (\bibinfo {year} {2009})}\BibitemShut {NoStop}%
\bibitem [{Note1()}]{Note1}%
  \BibitemOpen
  \bibinfo {note} {This is related to the flat dispersion of the C-shape mode
  at $\theta = \pi /4$ which leads to a high spectral density and consequently
  to modes very concentrated at few defect unit-cells.}\BibitemShut {Stop}%
\bibitem [{\citenamefont {Meenehan}\ \emph {et~al.}(2015)\citenamefont
  {Meenehan}, \citenamefont {Cohen}, \citenamefont {MacCabe}, \citenamefont
  {Marsili}, \citenamefont {Shaw},\ and\ \citenamefont
  {Painter}}]{PhysRevX.5.041002}%
  \BibitemOpen
  \bibfield  {author} {\bibinfo {author} {\bibfnamefont {S.~M.}\ \bibnamefont
  {Meenehan}}, \bibinfo {author} {\bibfnamefont {J.~D.}\ \bibnamefont {Cohen}},
  \bibinfo {author} {\bibfnamefont {G.~S.}\ \bibnamefont {MacCabe}}, \bibinfo
  {author} {\bibfnamefont {F.}~\bibnamefont {Marsili}}, \bibinfo {author}
  {\bibfnamefont {M.~D.}\ \bibnamefont {Shaw}}, \ and\ \bibinfo {author}
  {\bibfnamefont {O.}~\bibnamefont {Painter}},\ }\href {\doibase
  10.1103/PhysRevX.5.041002} {\bibfield  {journal} {\bibinfo  {journal} {Phys.
  Rev. X}\ }\textbf {\bibinfo {volume} {5}},\ \bibinfo {pages} {041002}
  (\bibinfo {year} {2015})}\BibitemShut {NoStop}%
\bibitem [{\citenamefont {Stockill}\ \emph {et~al.}(2022)\citenamefont
  {Stockill}, \citenamefont {Forsch}, \citenamefont {Hijazi}, \citenamefont
  {Beaudoin}, \citenamefont {Pantzas}, \citenamefont {Sagnes}, \citenamefont
  {Braive},\ and\ \citenamefont {Gr{\"o}blacher}}]{stockill2022ultra}%
  \BibitemOpen
  \bibfield  {author} {\bibinfo {author} {\bibfnamefont {R.}~\bibnamefont
  {Stockill}}, \bibinfo {author} {\bibfnamefont {M.}~\bibnamefont {Forsch}},
  \bibinfo {author} {\bibfnamefont {F.}~\bibnamefont {Hijazi}}, \bibinfo
  {author} {\bibfnamefont {G.}~\bibnamefont {Beaudoin}}, \bibinfo {author}
  {\bibfnamefont {K.}~\bibnamefont {Pantzas}}, \bibinfo {author} {\bibfnamefont
  {I.}~\bibnamefont {Sagnes}}, \bibinfo {author} {\bibfnamefont
  {R.}~\bibnamefont {Braive}}, \ and\ \bibinfo {author} {\bibfnamefont
  {S.}~\bibnamefont {Gr{\"o}blacher}},\ }\href@noop {} {\bibfield  {journal}
  {\bibinfo  {journal} {Nature Communications}\ }\textbf {\bibinfo {volume}
  {13}},\ \bibinfo {pages} {1} (\bibinfo {year} {2022})}\BibitemShut {NoStop}%
\bibitem [{\citenamefont {Schneider}\ \emph {et~al.}(2019)\citenamefont
  {Schneider}, \citenamefont {Baumgartner}, \citenamefont {Hönl},
  \citenamefont {Welter}, \citenamefont {Hahn}, \citenamefont {Wilson},
  \citenamefont {Czornomaz},\ and\ \citenamefont {Seidler}}]{Schneider:19}%
  \BibitemOpen
  \bibfield  {author} {\bibinfo {author} {\bibfnamefont {K.}~\bibnamefont
  {Schneider}}, \bibinfo {author} {\bibfnamefont {Y.}~\bibnamefont
  {Baumgartner}}, \bibinfo {author} {\bibfnamefont {S.}~\bibnamefont {Hönl}},
  \bibinfo {author} {\bibfnamefont {P.}~\bibnamefont {Welter}}, \bibinfo
  {author} {\bibfnamefont {H.}~\bibnamefont {Hahn}}, \bibinfo {author}
  {\bibfnamefont {D.~J.}\ \bibnamefont {Wilson}}, \bibinfo {author}
  {\bibfnamefont {L.}~\bibnamefont {Czornomaz}}, \ and\ \bibinfo {author}
  {\bibfnamefont {P.}~\bibnamefont {Seidler}},\ }\href {\doibase
  10.1364/OPTICA.6.000577} {\bibfield  {journal} {\bibinfo  {journal} {Optica}\
  }\textbf {\bibinfo {volume} {6}},\ \bibinfo {pages} {577} (\bibinfo {year}
  {2019})}\BibitemShut {NoStop}%
\bibitem [{\citenamefont {Kersul}\ \emph {et~al.}(2022)\citenamefont {Kersul},
  \citenamefont {Benevides}, \citenamefont {Moraes}, \citenamefont {de~Aguiar},
  \citenamefont {Wallucks}, \citenamefont {Gr{\"o}blacher}, \citenamefont
  {Wiederhecker},\ and\ \citenamefont {Alegre}}]{zenodo_cshape}%
  \BibitemOpen
  \bibfield  {author} {\bibinfo {author} {\bibfnamefont {C.~M.}\ \bibnamefont
  {Kersul}}, \bibinfo {author} {\bibfnamefont {R.}~\bibnamefont {Benevides}},
  \bibinfo {author} {\bibfnamefont {F.}~\bibnamefont {Moraes}}, \bibinfo
  {author} {\bibfnamefont {G.~H.~M.}\ \bibnamefont {de~Aguiar}}, \bibinfo
  {author} {\bibfnamefont {A.}~\bibnamefont {Wallucks}}, \bibinfo {author}
  {\bibfnamefont {S.}~\bibnamefont {Gr{\"o}blacher}}, \bibinfo {author}
  {\bibfnamefont {G.~S.}\ \bibnamefont {Wiederhecker}}, \ and\ \bibinfo
  {author} {\bibfnamefont {T.~P.~M.}\ \bibnamefont {Alegre}},\ }\href {\doibase
  10.5281/zenodo.7249807} {\enquote {\bibinfo {title} {{Data and simulation
  files for: "Silicon anisotropy in a bi-dimensional optomechanical
  cavity"}},}\ } (\bibinfo {year} {2022})\BibitemShut {NoStop}%
\end{thebibliography}%


\begin{thebibliography}{4}%
\makeatletter
\providecommand \@ifxundefined [1]{%
 \@ifx{#1\undefined}
}%
\providecommand \@ifnum [1]{%
 \ifnum #1\expandafter \@firstoftwo
 \else \expandafter \@secondoftwo
 \fi
}%
\providecommand \@ifx [1]{%
 \ifx #1\expandafter \@firstoftwo
 \else \expandafter \@secondoftwo
 \fi
}%
\providecommand \natexlab [1]{#1}%
\providecommand \enquote  [1]{``#1''}%
\providecommand \bibnamefont  [1]{#1}%
\providecommand \bibfnamefont [1]{#1}%
\providecommand \citenamefont [1]{#1}%
\providecommand \href@noop [0]{\@secondoftwo}%
\providecommand \href [0]{\begingroup \@sanitize@url \@href}%
\providecommand \@href[1]{\@@startlink{#1}\@@href}%
\providecommand \@@href[1]{\endgroup#1\@@endlink}%
\providecommand \@sanitize@url [0]{\catcode `\\12\catcode `\$12\catcode
  `\&12\catcode `\#12\catcode `\^12\catcode `\_12\catcode `\%12\relax}%
\providecommand \@@startlink[1]{}%
\providecommand \@@endlink[0]{}%
\providecommand \url  [0]{\begingroup\@sanitize@url \@url }%
\providecommand \@url [1]{\endgroup\@href {#1}{\urlprefix }}%
\providecommand \urlprefix  [0]{URL }%
\providecommand \Eprint [0]{\href }%
\providecommand \doibase [0]{http://dx.doi.org/}%
\providecommand \selectlanguage [0]{\@gobble}%
\providecommand \bibinfo  [0]{\@secondoftwo}%
\providecommand \bibfield  [0]{\@secondoftwo}%
\providecommand \translation [1]{[#1]}%
\providecommand \BibitemOpen [0]{}%
\providecommand \bibitemStop [0]{}%
\providecommand \bibitemNoStop [0]{.\EOS\space}%
\providecommand \EOS [0]{\spacefactor3000\relax}%
\providecommand \BibitemShut  [1]{\csname bibitem#1\endcsname}%
\let\auto@bib@innerbib\@empty
%</preamble>
\bibitem [{\citenamefont {Safavi-Naeini}\ and\ \citenamefont
  {Painter}(2010)}]{safavinaeini2010}%
  \BibitemOpen
  \bibfield  {author} {\bibinfo {author} {\bibfnamefont {A.~H.}\ \bibnamefont
  {Safavi-Naeini}}\ and\ \bibinfo {author} {\bibfnamefont {O.}~\bibnamefont
  {Painter}},\ }\bibfield  {title} {\enquote {\bibinfo {title} {{Design of
  optomechanical cavities and waveguides on a simultaneous bandgap
  phononic-photonic crystal slab}},}\ }\href {\doibase 10.1364/OE.18.014926}
  {\bibfield  {journal} {\bibinfo  {journal} {Optics Express}\ }\textbf
  {\bibinfo {volume} {18}},\ \bibinfo {pages} {14926} (\bibinfo {year}
  {2010})}\BibitemShut {NoStop}%
\bibitem [{\citenamefont {Safavi-Naeini}\ and\ \citenamefont
  {Painter}(2014)}]{safavi2014optomechanical}%
  \BibitemOpen
  \bibfield  {author} {\bibinfo {author} {\bibfnamefont {A.~H.}\ \bibnamefont
  {Safavi-Naeini}}\ and\ \bibinfo {author} {\bibfnamefont {O.}~\bibnamefont
  {Painter}},\ }\bibfield  {title} {\enquote {\bibinfo {title} {Optomechanical
  crystal devices},}\ }in\ \href@noop {} {\emph {\bibinfo {booktitle} {Cavity
  Optomechanics}}}\ (\bibinfo  {publisher} {Springer},\ \bibinfo {year}
  {2014})\ pp.\ \bibinfo {pages} {195--231}\BibitemShut {NoStop}%
\bibitem [{\citenamefont {Safavi-Naeini}\ \emph {et~al.}(2014)\citenamefont
  {Safavi-Naeini}, \citenamefont {Hill}, \citenamefont {Meenehan},
  \citenamefont {Chan}, \citenamefont {Gr\"oblacher},\ and\ \citenamefont
  {Painter}}]{PhysRevLett.112.153603}%
  \BibitemOpen
  \bibfield  {author} {\bibinfo {author} {\bibfnamefont {A.~H.}\ \bibnamefont
  {Safavi-Naeini}}, \bibinfo {author} {\bibfnamefont {J.~T.}\ \bibnamefont
  {Hill}}, \bibinfo {author} {\bibfnamefont {S.}~\bibnamefont {Meenehan}},
  \bibinfo {author} {\bibfnamefont {J.}~\bibnamefont {Chan}}, \bibinfo {author}
  {\bibfnamefont {S.}~\bibnamefont {Gr\"oblacher}}, \ and\ \bibinfo {author}
  {\bibfnamefont {O.}~\bibnamefont {Painter}},\ }\bibfield  {title} {\enquote
  {\bibinfo {title} {Two-dimensional phononic-photonic band gap optomechanical
  crystal cavity},}\ }\href {\doibase 10.1103/PhysRevLett.112.153603}
  {\bibfield  {journal} {\bibinfo  {journal} {Phys. Rev. Lett.}\ }\textbf
  {\bibinfo {volume} {112}},\ \bibinfo {pages} {153603} (\bibinfo {year}
  {2014})}\BibitemShut {NoStop}%
\bibitem [{\citenamefont {Gorodetksy}\ \emph {et~al.}(2010)\citenamefont
  {Gorodetksy}, \citenamefont {Schliesser}, \citenamefont {Anetsberger},
  \citenamefont {Deleglise},\ and\ \citenamefont {Kippenberg}}]{Gorodetksy:10}%
  \BibitemOpen
  \bibfield  {author} {\bibinfo {author} {\bibfnamefont {M.~L.}\ \bibnamefont
  {Gorodetksy}}, \bibinfo {author} {\bibfnamefont {A.}~\bibnamefont
  {Schliesser}}, \bibinfo {author} {\bibfnamefont {G.}~\bibnamefont
  {Anetsberger}}, \bibinfo {author} {\bibfnamefont {S.}~\bibnamefont
  {Deleglise}}, \ and\ \bibinfo {author} {\bibfnamefont {T.~J.}\ \bibnamefont
  {Kippenberg}},\ }\bibfield  {title} {\enquote {\bibinfo {title}
  {Determination of the vacuum optomechanical coupling rate using frequency
  noise calibration},}\ }\href {\doibase 10.1364/OE.18.023236} {\bibfield
  {journal} {\bibinfo  {journal} {Opt. Express}\ }\textbf {\bibinfo {volume}
  {18}},\ \bibinfo {pages} {23236--23246} (\bibinfo {year} {2010})}\BibitemShut
  {NoStop}%
\end{thebibliography}%

\end{document}

% \preprint{AIP/123-QED}

\title{Supplementary Material: Silicon anisotropy in a bi-dimensional optomechanical cavity}
% Force line breaks with \\

\author{Cau\^e M. Kersul}
\altaffiliation{These two authors contributed equally} 
\affiliation{Photonics Research Center, University of Campinas, Campinas, SP, Brazil}
\affiliation{Applied Physics Department, Gleb Wataghin Physics Institute, University of Campinas, Campinas, SP, Brazil}

\author{Rodrigo Benevides}
%\altaffiliation[Now at ]{Department of Physics, ETH Zürich, 8093 Zürich, Switzerland} 
\altaffiliation{These two authors contributed equally} 
\affiliation{Photonics Research Center, University of Campinas, Campinas, SP, Brazil}
\affiliation{Applied Physics Department, Gleb Wataghin Physics Institute, University of Campinas, Campinas, SP, Brazil}
\affiliation{Kavli Institute of Nanoscience, Department of Quantum Nanoscience, Delft University of Technology, Delft, The Netherlands}

\author{Fl\'avio Moraes}
\affiliation{Photonics Research Center, University of Campinas, Campinas, SP, Brazil}
\affiliation{Applied Physics Department, Gleb Wataghin Physics Institute, University of Campinas, Campinas, SP, Brazil}

\author{Gabriel H. M. de Aguiar}
\affiliation{Photonics Research Center, University of Campinas, Campinas, SP, Brazil}
\affiliation{Applied Physics Department, Gleb Wataghin Physics Institute, University of Campinas, Campinas, SP, Brazil}

\author{Andreas Wallucks}
\affiliation{Kavli Institute of Nanoscience, Department of Quantum Nanoscience, Delft University of Technology, Delft, The Netherlands}

\author{Simon Gr\"{o}blacher}
\affiliation{Kavli Institute of Nanoscience, Department of Quantum Nanoscience, Delft University of Technology, Delft, The Netherlands}

\author{Gustavo S. Wiederhecker}
\affiliation{Photonics Research Center, University of Campinas, Campinas, SP, Brazil}
\affiliation{Quantum Electronics Department, Gleb Wataghin Physics Institute, University of Campinas, Campinas, SP, Brazil}

\author{Thiago P. Mayer Alegre}
\email{alegre@unicamp.br}
\affiliation{Photonics Research Center, University of Campinas, Campinas, SP, Brazil}
\affiliation{Applied Physics Department, Gleb Wataghin Physics Institute, University of Campinas, Campinas, SP, Brazil}

\maketitle

\section{C-shape geometry}

\subsection{Design of the fabricated devices}

\begin{figure}[htp]
\centering
\includegraphics{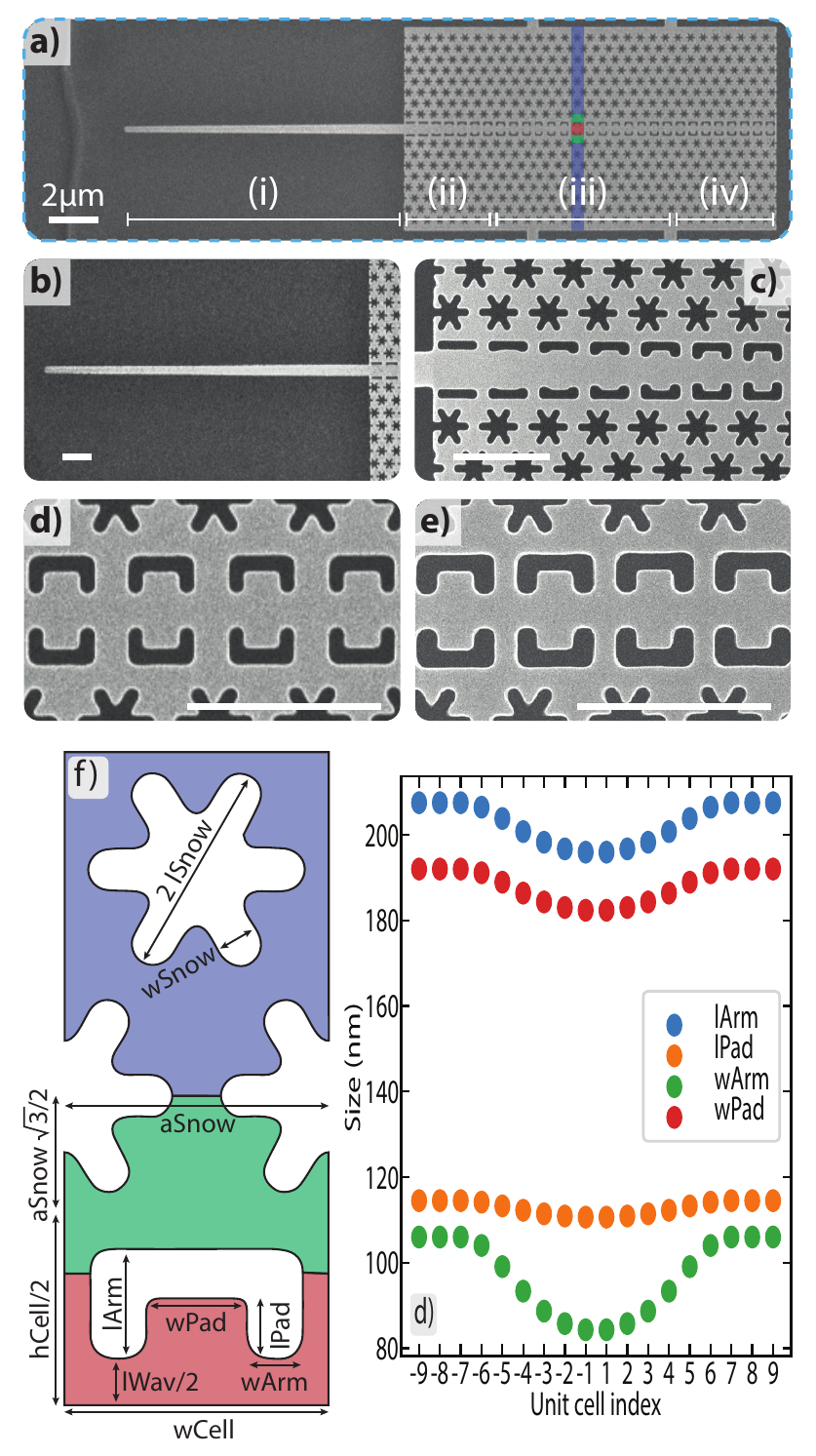}
\caption{\textbf{a)} SEM image of a single device indicating its different sections. \textbf{b)} Zoom of section (i), the tapered integrated waveguide. \textbf{c)} Zoom of section (ii) the transition region. \textbf{d)} Zoom of section (iii), the central defect unit-cells. \textbf{e)} Zoom of section (iv), the mirror unit-cells. The white bars in \textbf{b)-e)} indicate a \SI{1}{\micro\meter} scale. \textbf{f)} Scheme indicating the different parameters of the geometry. \textbf{g)} Parameter variation across the unit-cells of the simulated structure}
\label{fig1_supp}
\end{figure}

Our device is composed of a quasi-1D optomechanical crystal cavity surrounded by quasi-2D optomechanical crystal shields. The quasi-1D optomechanical crystal cavity is formed by two lines of C-shaped holes facing each other, while the quasi-2D optomechanical crystal shields are formed by a triangular lattice of snowflake-shaped holes, as shown in Fig.\ref{fig1_supp}. Such holes are defined in a \SI{220}{nm} thick suspended slab of silicon. The fabricated devices are divided in 4 regions: (i) the tapered integrated waveguide, (ii) a transition region where the C-shape format is slowly introduced, (iii) the defect region and (iv) the final mirror region.

\paragraph{(i) Tapered integrated waveguide}

The width of the tapered integrated waveguide is varied from \SI{200}{nm} to \SI{360}{nm} along \SI{10}{\micro\meter} in order to adiabatically couple light from a tapered optical fiber to the integrated waveguide, as shown in Fig.~\ref{fig1_supp}~\textbf{b}.

\paragraph{(ii) Optomechanical crystal transition region}

In order to mitigate optical reflection the C-shape is slowly introduced to the optomechanical crystal structure, as shown in Fig.~\ref{fig1_supp}~\textbf{c}. At the end of this region the unit-cell has the same shape as the unit-cells at the mirror region. 

\paragraph{(iii) and (iv) Optomechanical crystal defect and mirror regions}

The transition region is followed by the defect region, in which the unit-cell shape is gradually varied from the mirror unit-cell shape, shown in Fig.~\ref{fig1_supp}~\textbf{d}, to the central defect unit-cell shape, shown in Fig.~\ref{fig1_supp}~\textbf{e}. Finally, going from the center of the defect region to the mirror region the unit-cell changes from the central defect shape to the mirror shape. The total length of the defect region is 14 cells, while the mirror region is composed of 8 cells.   

In Fig.~\ref{fig1_supp}~\textbf{f} we label the different geometrical parameters related to the shape and to the spacing of the C-shape and snowflake holes. In our design, the defect is created by decreasing the parameters lArm, wArm, lPad and wPad, from their maxima at the mirror unit-cells to minima at the central defect cells, following the expression:

\begin{equation}
    \mathrm{Par(n)} = \mathrm{Par_{min}} + (\mathrm{Par_{max}}-\mathrm{Par_{min}}).\;e^{\mathrm{\frac{9 (|n|-n_{d})^2}{2n_{d}^2}}},
\end{equation}

%Exp\left(\frac{9 (|n|-n_{d})^2}{2n_{d}^2}\right)

\noindent where $\rm Par(n)$ is the the parameter of the unit cell with index $\rm n$, $\rm Par_{min}$ and $\rm Par_{max}$ are the minimum and the maximum values of the parameter, and $\rm n_d$ is half the number of defect unit-cells. In Fig.~\ref{fig1_supp}~\textbf{g} we plot the the C-shape parameters as a function of the unit cell index, the defect region corresponds to the unit cells between $\rm |n|\leq 7$, while the mirror region corresponds to $\rm |n|>7$. 

The values of the fixed and variable parameters used in the simulations of the whole device discussed at Fig.~\textbf{3} are presented in Tab.~\ref{supp:tab1}. These values are based on the SEM images of the fabricated devices. In order to speed up the simulations, only the mirror and the defect regions are simulated. This is done by imposing a $\pi$ cyclic symmetry around the center of the defect region. For more details we refer the reader to the .mph files in the Zenodo.

\begin{table}[h!]
\centering
\begin{tabular}{c|c|c}
\hline
Parameter & Value (\SI{}{nm}) & Description \\ \hline \hline
lWav & 175.0 & Vertical separation between two C-shape holes \\ \hline
wCell & 512.5 & Width of the unit cell \\ \hline
n$_{\rm{d}}$ & 7 & Half the number of defects \\ \hline
lArm$_{\rm{max}}$ & 207.5 & Maximum of length of the lateral C-shape arms \\ \hline
lArm$_{\rm{min}}$ & 195.5 & Minimum of length of the lateral C-shape arms \\ \hline
wArm$_{\rm{max}}$ & 106.0 & Maximum of width of the lateral C-shape arms \\ \hline
wArm$_{\rm{min}}$ &  83.5 & Minimum of width of the lateral C-shape arms \\ \hline
lPad$_{\rm{max}}$ & 110.5 & Maximum of length of the C-shape pad \\ \hline
lPad$_{\rm{min}}$ & 114.5 & Minimum of length of the C-shape pad \\ \hline
wPad$_{\rm{max}}$ & 192.0 & Maximum of width of the C-shape pad \\ \hline
wPad$_{\rm{min}}$ & 182.0 & Minimum of width of the C-shape pad \\ \hline
aSnow & 500.0 & Separation between snowflake holes\\ \hline
lSnow & 205.0 & Length of the snowflake arms\\ \hline
wSnow &  82.0 & Width of the snowflake arms\\ \hline
\end{tabular}
\caption{Value of the geometrical parameters used in the simulation.}
\label{supp:tab1}
\end{table}

\subsection{Microfabrication process}
The fabrication steps of the devices follow a basic CMOS-compatible top-down approach. We start with a die of an SOI (silicon-on-insulator) wafer over which we spin CSAR-09 electroresist at 2000 rpm for 1 minute. We write the desired geometries on the resist with the help of a $\SI{100}{kV}$ electron beam lithography equipment. After it, we dip the chip in a pentyl-acetate solution for 1 minute, developing the patterns on the resist. We transfer this pattern to the silicon layer with the help of an SF$_6$+O$_2$ plasma etching at cryogenic temperature, to obtain smooth walls. Finally, we clean the resist with an N,N-dimethylformamide, and a piranha process (H$_2$SO$_4$:H$_2$O$_2$ - 3:1) for 10 minutes and proceed to the release process of the crystal, with a solution of HF (hydrofluoric acid) for 3 minutes.

\section{Band structure}
As mentioned in the main text the C-shape geometry discussed here is a way to combine high optical and mechanical quality factors and high optomechanical couplings from 1D geometries, such as the nanobeam, with the faster thermal response of 2D optomechanical crystals. Both the snowflake 2D optomechanical crystal and the C-shape structure are known to present full optical and mechanical bandgaps \cite{safavinaeini2010}.

The optical mode of interest is concentrated inside the central C-shaped holes, with the optical field in the $y$ direction, see Fig.~\textbf{1 c}, much like in a slot waveguide mode, the frequency of such modes is strongly dependent on the width of the airgap (lArm-lPad). The mechanical mode of interest is the breathing mode of the rectangular paddles in between the C-shaped holes, see Fig.~\textbf{1 d}, whose frequency is inversely proportional to the length of the pads (lPad). This geometry leads to a strong moving boundary optomechanical coupling, as the width of the C-shaped holes is modulated by the mechanical motion of the paddles. The photoelastic coupling is also important as the dominant component of the mechanical strain, $\sigma_{yy}$, is coupled strongly with the dominant component of the electric energy $|E_y|^2$ through $p_{11} = -0.094$, the highest component of silicon photoelastic tensor.

\begin{figure*}[htp]
\centering
\includegraphics{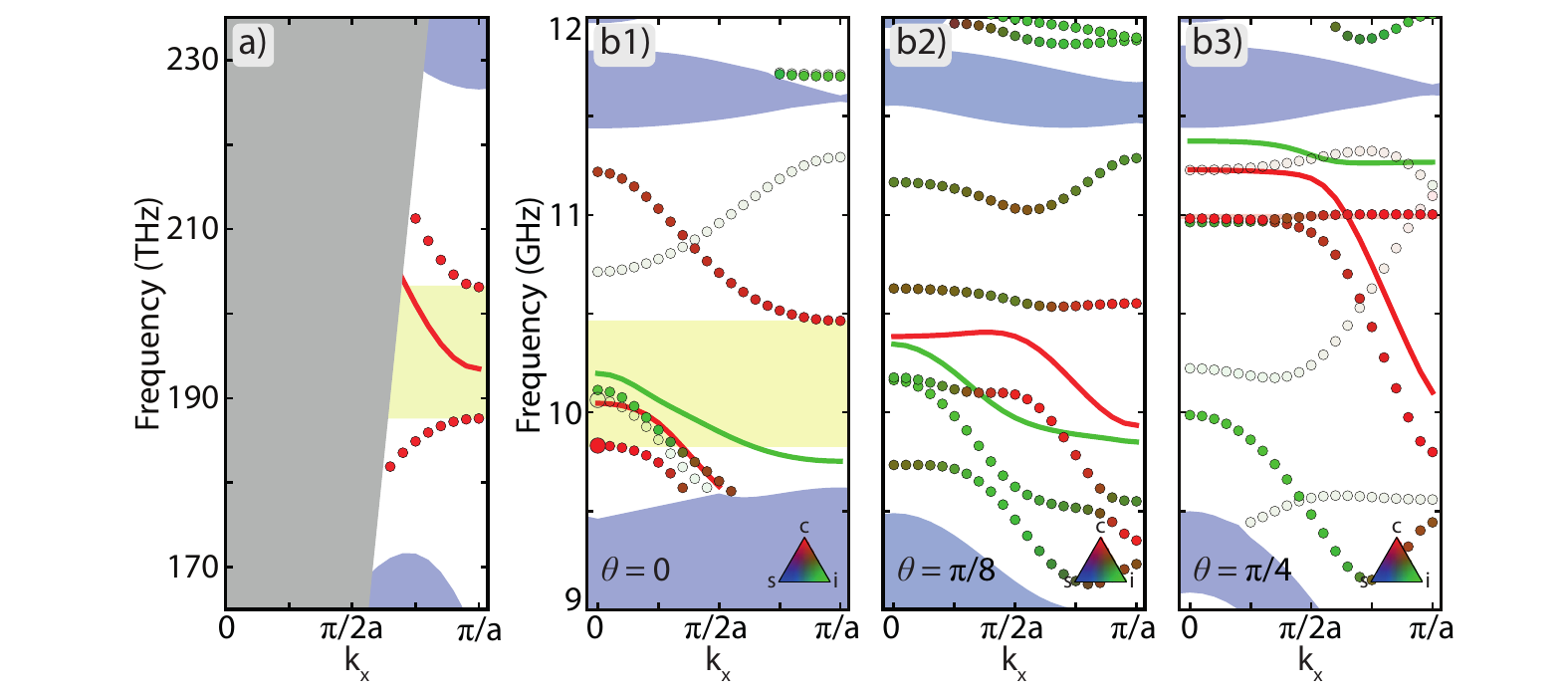}
\caption{\textbf{a)} Optical band structure, the line indicates the confined mode of the defect unit-cell. \textbf{b)} Mechanical band structures for $\theta = 0, \pi/8 \;\mathrm{and}\; \pi/4$, respectively. The solid lines indicate different modes of the defect unit-cell, the red ones indicate modes predominantly at the C-shape region, while the green one indicate modes closer to the interface.} 
\label{fig2_supp}
\end{figure*}

In Fig.~\ref{fig2_supp}~\textbf{a} we show the optical band structure of the mirror unit-cell shown in Fig.~\textbf{2 a}. The optical cone of light is shaded in gray, in this region, modes are no longer confined within the silicon device layer. In blue we present the snowflake bands, these modes are vertically confined within the snowflake region, but are unconfined in the direction of the plane. The C-shape modes are designed to lie within the snowflake band gap, preventing them to loose energy through in-plane radiation. These modes define the band gap for the full hybrid structure, shaded in yellow in Fig.~\ref{fig2_supp}~\textbf{a}. The crystal defect for the optical modes is designed by decreasing the width of the C-shape airgap. This decreases the frequency of the optical mode, confining the upper band of C-shape optical modes within the band gap. As pointed in \onlinecite{safavi2014optomechanical} this optical mode is confined near $k_x = \pi/a$, as modes near $k_x = 0$ are within the light cone.

As discussed in the main text the  mechanical band structure depends on the angle $\theta$ between the device and the [100] crystalline direction of silicon. For $\theta = 0$ the mechanical band structure is similar to the optical one, we can define a C-shape band gap within the larger snowflake band gap. Nevertheless it has two fundamental differences with respect to the optical band structure. Firstly the mechanical modes cannot propagate in vacuum in such a way that we do not have an equivalent to the light cone, allowing confinement near $k_x = 0$, and secondly, interface modes, located in between the snowflake and the C-shape modes, cross the C-shape band gap, as seen in Fig.~\ref{fig2_supp} \textbf{b1}. The defect in such structure is created by decreasing the length of the c-shape paddle, increasing the frequency of the mechanical breathing mode near $k_x = 0$, represented by the solid red line in Fig.~\ref{fig2_supp} b1). This mode is confined in the smaller band gap near $k_x = 0$ in between the c-shape and the interface mode. 

%The proximity to the interface mode leads to spurious couplings when we consider angles different from $\theta = 0$, as seen in Fig.~\textbf{3 b} of the main text, decreasing the overall optomechanical coupling.

For $\theta = \pi/4$ the mechanical band structure changes considerably, instead of a C-shape band gap, we have a band linking the c-shape modes from $k_x = 0$ to $k_x = \pi/a$, represented by the red shaded area in Fig.~\ref{fig2_supp} b3). This band is very close in frequency to another interface band. Both the c-shape and the interface bands are flat, in special near $k_x = 0$, leading to extremely confined c-shape and interface modes \cite{PhysRevLett.112.153603}. When we consider defect unit cells the frequency of such modes increases leading to an anti-crossing between them two. The presence of multiple C-shape and interface modes very close in frequency and well confined in the same space makes the device extremely susceptible to spurious couplings between such c-shape and interface modes due to fabrication irregularities. This can be seem in Fig.~\textbf{3 b}, where we perceive the high number of modes around the best confined c-shape mode.

% As discussed in \cite{amir}, optomechanical crystals can be created by designing a defect on the unit cell structure with optical and mechanical modes within the bandgaps of the mirror bandstructure. Optical modes are confined around $k_x = \pi/a$, as modes near $k_x = 0$ are within the light cone. Mechanical modes are confined near $k_x = 0$ in order to assure that neighboring unit cells add constructively to the overall optomechanical coupling.

\section{Optomechanical measurements}

\begin{figure*}[htp]
\centering
\includegraphics{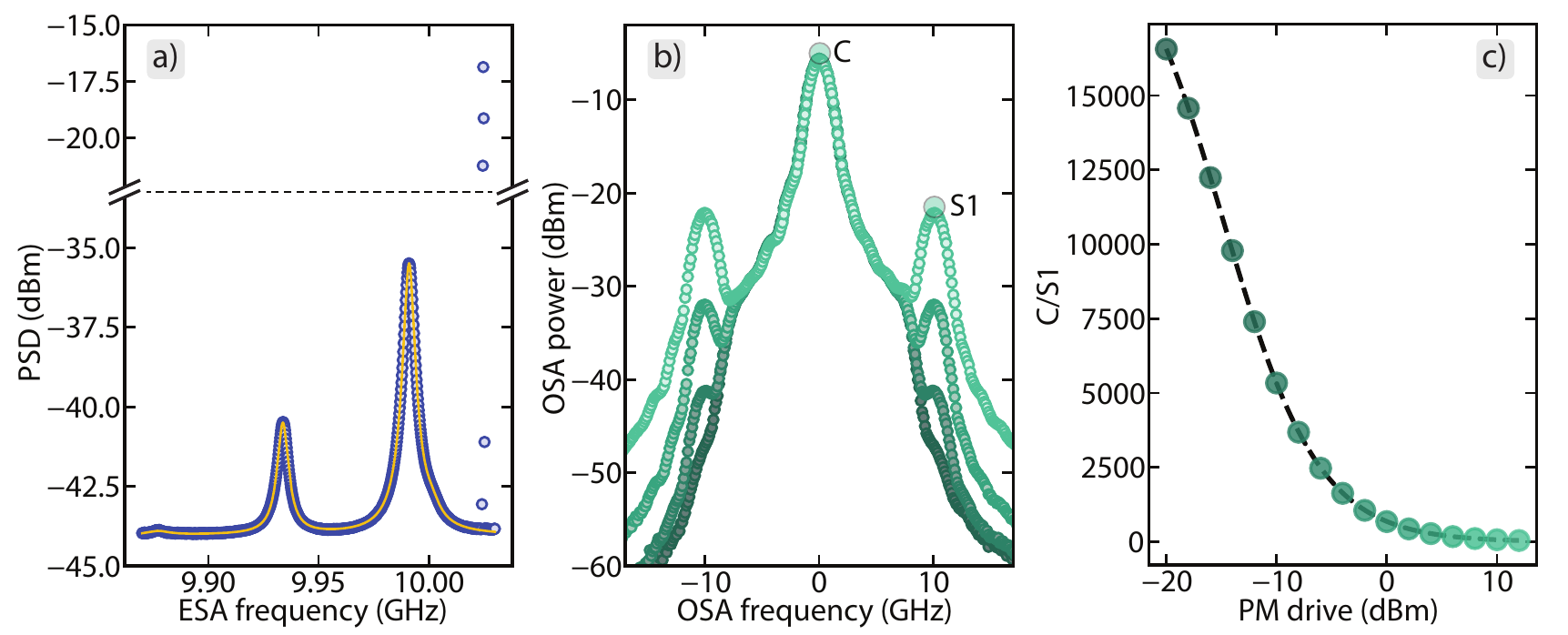}
\caption{\textbf{a)} ESA spectrum measured showing the optomechanical and the calibrated phase modulation signals. The solid yellow line presents the fit of the spectrum considering 4 modes. The $y-$axis is cut in order to better show the peak of the phase modulation signal. \textbf{b)} Example of optical spectrum analyzer (OSA) spectra of the modulated laser at different PM driving powers, showing the central band (C) surrounded by the phase modulation sidebands (S1). \textbf{c)} Ratio between the central and the sideband as a function of the drive power on the PM, the dots are experimental data, while the dashed line is the fit used to extract $V_{\pi}$.} 
\label{fig3_supp}
\end{figure*}

In order to characterize the optomechanical coupling we have used the method discussed in ~\onlinecite{Gorodetksy:10}. A phase modulator, "PM" in Fig.~\textbf{1 e}, introduces a calibrated phase modulation sideband to the laser feeding the optomechanical cavity. This modulation is transduced from the phase to the amplitude quadrature by the dispersive nature of the optical reflection near an optical resonance frequency (Fig.~\textbf{1 f}). The same transduction process happens to the sidebands introduced by the mechanical modes. Finally, both these signals appear in the power spectral density (PSD) of the reflection signal, as shown in Fig.~\ref{fig3_supp} a), measured in a electric spectrum analyzer (ESA).

As the frequency of the calibrated phase modulation, $\Omega_{\textrm{mod}}/2\pi$, is very close to the frequency of the mechanical modes, $\Omega_{\textrm{m}}/2\pi$, the transduction factor from phase to amplitude is approximately the same for both. Comparing the PSD peaks for the optomechanical, $\mathrm{PSD_{m}}$, and for the calibrated phase, $\mathrm{PSD}_{\phi}$, modulations we can directly calculate the ratio between the respective phase modulations, bypassing the need to condider the transduction factor:

%    \frac{\mathrm{PSD}_{m}}{\mathrm{PSD}_{\phi}} = \frac{\mathrm{ENBW}}{b^2\Omega^2} g_0^2 \frac{S_{xx}^{\mathrm{peak}}}{x_{\mathrm{zpf}}^2},

\begin{equation}
    \frac{\mathrm{PSD_m}}{\mathrm{PSD}_{\phi}} = \mathrm{ENBW} \frac{ \; g_0^2 \; S_{xx}^{\mathrm{peak}}/x_{\mathrm{zpf}}^2}{b^2\Omega^2},
    \label{eq1}
\end{equation}
where $g_0$ is the optomechanical coupling rate, $\Omega$ is the mechanical mode frequency, $S_{xx}^{\mathrm{peak}}$ is its displacement power spectral density, $x_{\mathrm{zpf}}$ is its zero point fluctuation displacement, $\mathrm{ENBW}$ is the effective noise bandwidth of the ESA filter and $b$ is the calibrated phase modulation amplitude. As the calibrated phase modulation presents a bandwidth much narrower than $\mathrm{ENBW}$ and the optomechanical signal presents a linewidth ($\Gamma$) much larger than $\mathrm{ENBW}$, the latter must be multiplied by $\mathrm{ENBW}$ in order to properly compare them. 

For the case of Brownian motion at a temperature $T$, one can calculate:
\begin{equation}
    \frac{S_{xx}^{\mathrm{peak}}}{ x_{\mathrm{zpf}}^2} = \frac{4 k_\mathrm{b}T}{\hbar}\frac{\Gamma\Omega}{\Gamma^2\Omega^2 - \Gamma^4/4},
    \label{eq2}
\end{equation}
where $k_\mathrm{b}$ is the Boltzmann constant, $\hbar$ is the reduced Plank constant. 

Replacing Eq.~\ref{eq2} in Eq.~\ref{eq1} we get:

\begin{equation}
    g_0^2 = \frac{\mathrm{PSD}_{m}}{\mathrm{PSD}_{\phi}} \left(\Gamma\Omega - \frac{\Gamma^3}{4\Omega}\right) \frac{\hbar b^2 \Omega^2}{4k_b T \; \mathrm{ENBW}}.
\end{equation}
Where $g_0$ is written as a function of measured parameters and of fundamental constants. The parameters from the optomechanical signal are fitted considering multiple modes using the function:
\begin{equation}
    PSD_m(f) = PSD_{0} + \sum_i PSD_{m,i}^{\mathrm{peak}} \frac{\Omega_i^2\Gamma_i^2-\Gamma_i^4/4}{(\Omega_i^2-(f/2\pi)^2)^2+\Gamma_i^2(f/2\pi)^2},
\end{equation}
where $PSD_{0}$ is the background PSD signal.

The PM is calibrated by measuring the modulated laser spectrum, Fig.~\ref{fig3_supp}~\textbf{b} as a function of the PM drive power. In Fig.\ref{fig3_supp}~\textbf{c} the ratio between the central band, (C), and the first sideband, (S1), is fitted according to the function:
\begin{equation}
    \frac{C}{S1} = \frac{J_0(b)^2}{J_1(b)^2 + B}, \textrm{    with,   } b = \frac{\pi}{V_{\pi}}\sqrt{2RP_{\textrm{PM}}}
\end{equation}
where $J_n$ is the $n-$th  Bessel function of the first kind, $B$ is a fitting parameter related to the noise floor for S1, $V_{\pi}$ is the fitting parameter used to characterize the PM, $R = \SI{50}{\ohm}$ is the PM internal impedance and $P_{\textrm{PM}}$ is the electric power driving the PM.

\subsection{Sideband asymmetry measurements}

The low-temperature experiment is performed in a dilution refrigerator base plate, kept at $T\sim 20 mK$. The relationship between the scattering rates of optical pulses incident in the device at detunings $\Delta=\pm \Omega_m$ allows us to characterize the temperature of the mechanical mode and, consequently, its phonon occupation level. To avoid heating the cavity, we use low-power pulses (below $15$~nW), leading to low scattering probabilities and a low success rate of the event. Such a low power is also useful when one wants to work on a mechanical qubit basis, allowing one to disregard high occupation level in the mechanical mode. 

Pump photons are filtered from the pulse after the cavity using a system of narrow bandwidth filters ($\sim 40 $~MHz), which allows us to guarantee that most of the detected photons are either Stokes or anti-Stokes generated photons. The detection of the scattered photons is done using superconducting nanowire single-photon detectors.

To increase the measurement precision, we take into consideration the number of pump-leaked photons that pass through the filters and reach the detectors. We find this number by detuning the laser far away from the cavity and using the equivalent optical power. We also include the dark count photons, which are intrinsic noise photons in the experiment and are independent of our optical pulses. Both numbers are subtracted from our experimental data to perform the thermometry measurement. Data and processing codes are found in the Zenodo files. 

\bibliography{cshape-ref}% Produces the bibliography via BibTeX.